\documentclass[12pt, letterpaper]{extarticle}
\usepackage{xcolor}
\usepackage{makecell}
\usepackage[hyphens]{url}
\usepackage{adjustbox}
\usepackage[flushleft]{threeparttable}
\usepackage{booktabs}
\usepackage{graphics}
\usepackage{helvet}
\usepackage{times}
\usepackage{comment}
\usepackage{placeins}
\usepackage{multirow}
\usepackage{rotating}
\usepackage{mathtools}
\usepackage{subfig}
\usepackage{arydshln}
\usepackage{longtable}
\usepackage{float}
\usepackage{multicol}
\usepackage{supertabular}
\usepackage{booktabs}
\usepackage{xcolor}
\usepackage{longtable}
\usepackage[framemethod=TikZ]{mdframed}
\usepackage{fullpage}
\usepackage[switch]{lineno}
\usepackage{amsmath}
\usepackage{amssymb}
\usepackage{rotating}
\usepackage{array}
\usepackage{mathtools}
\usepackage[ruled]{algorithm2e}
\usepackage{algorithmic}
\usepackage{bm}
\usepackage{breqn}
\usepackage{comment}
\usepackage{graphics}
\usepackage{graphicx}
\usepackage{latexsym}
\usepackage{mathrsfs}
\usepackage{morefloats}
\usepackage{nicefrac}
\usepackage{authblk}
\usepackage{ulem}
\usepackage{svg}
\usepackage{subfiles}
\usepackage{xr-hyper}
\usepackage[breaklinks]{hyperref}

\makeatletter
\newcommand*{\addFileDependency}[1]{
  \typeout{(#1)}
  \@addtofilelist{#1}
  \IfFileExists{#1}{}{\typeout{No file #1.}}
}
\makeatother

\graphicspath{{main_figures/}{../main_figures/}}

\definecolor{normgreen}{rgb}{0.0, 0.5, 0.0}

\title{China and the U.S.\ produce more impactful AI research when collaborating together}

\author[1*]{Bedoor AlShebli}
\author[1,2]{Shahan Ali Memon}
\author[3]{James A. Evans}
\author[4*]{Talal Rahwan}

\affil[1]{\normalsize 
Social Science Division, New York University Abu Dhabi, UAE.}

\affil[2]{\normalsize
Information School, University of Washington, WA, USA.}

\affil[3]{\normalsize Department of Sociology, University of Chicago, Chicago, IL, USA.}

\affil[4]{\normalsize
Science Division, New York University Abu Dhabi, UAE.}

\affil[*]{\footnotesize Joint corresponding authors. E-mails:\ bedoor@nyu.edu;\ talal.rahwan@nyu.edu}

\date{}

\begin{document} 
\maketitle

\begin{abstract}

Artificial Intelligence (AI) has become a disruptive technology, promising to grant a significant economic and strategic advantage to nations that harness its power. China, with its recent push towards AI adoption, is challenging the U.S.’s position as the global leader in this field. Given AI's massive potential, as well as the fierce geopolitical tensions between China and the U.S., several recent policies have been put in place to discourage AI scientists from migrating to, or collaborating with, the other nation. Nevertheless, the extent of talent migration and cross-border collaboration are not fully understood. Here, we analyze a dataset of over 350,000 AI scientists and 5,000,000 AI papers. We find that since 2000, China and the U.S.\ have led the field in terms of impact, novelty, productivity, and workforce. Most AI scientists who move to China come from the U.S., and most who move to the U.S. come from China, highlighting a notable bidirectional talent migration. Moreover, the vast majority of those moving in either direction have Asian ancestry. Upon moving, those scientists continue to collaborate frequently with those in the origin country. Although the number of collaborations between the two countries has increased since the dawn of the millennium, such collaborations continue to be relatively rare. A matching experiment reveals that the two countries have always been more impactful when collaborating than when each works without the other. These findings suggest that instead of suppressing cross-border migration and collaboration between the two nations, the science could benefit from promoting such activities.
\end{abstract}

\section*{Introduction}

Artificial Intelligence (AI) has become a disruptive technology with far-reaching economic, social, political and regulatory repercussions. Recent advances in robotics and automation continue to reshape local labor markets and the global employment landscape \cite{frey2017future, frank2018small, acemoglu2018race, frank2019toward, felten2021occupational, chen2021automation}. In healthcare, machine learning algorithms support the global response to pandemic outbreaks \cite{luengo2020artificial}, transforming the medical image analysis field \cite{kourou2015machine, hosny2018artificial}, and making drug discovery faster, cheaper, and more effective \cite{fleming2018artificial}. On the road, autonomous vehicles hold the promise of improving traffic flow, reducing pollution, and preventing traffic accidents that result from human error \cite{van2006impact, spieser2014toward, ecola2018road}. In decision making, algorithms are poised to address some of the major societal challenges of today, e.g., by reducing gender bias in hiring decisions \cite{erel2018could, mullainathan2019biased}, dropping crime rates with more informed bail \cite{kleinberg2018human} and enforcement decisions \cite{rotaru2022event}, and improving the way societies are governed \cite{rahwan2018society}. AI can even be helpful in the fight against corruption, with several governments and non-governmental organizations (NGOs) implementing AI-based anti-corruption tools that enable citizens to keep their bureaucratic officials in check \cite{kobis2022promise}. Nevertheless, despite its potential, the perils of AI are too consequential to ignore. Artificially intelligent systems may reduce human error, but they may also exacerbate discrimination against minorities by targeting disadvantaged groups or training on data that reflect systematic and persistent biases \cite{Daneshjou:2022, cowgill2020biased, barocas2016big, crawford2016there, saunders2016predictions, sweeney2013discrimination}. Moreover, malicious AI systems may disrupt peace and incite violence by spreading false information \cite{chesney2019deep, lazer2018science, allcott2017social} or increasing the threat of terrorism and autonomous weapons \cite{schmidt2021national, brundage2018malicious, scharre2016autonomous}.

Notwithstanding its perils, the technologies driven by AI are likely to underpin the security, prosperity, and welfare of the nations that harness them \cite{vinuesa2020role}. Given its potential to shape global competitiveness, the race for world leadership in AI adoption is intensifying globally, with countries developing national AI strategies in an effort to guide and foster its deployment through targeted investments and strategic collaborations \cite{AIindex2021}. China’s \textit{New Generation AI Development Plan} \cite{ChinaAIPlan} manifests its commitment towards making AI the driving force behind its industrial and economic transformation by 2025, and making China one of the world’s primary AI innovation centers by 2030 \cite{wu2020towards}. According to the latest report by the U.S.'s National Security Commission on AI, the U.S.\ could lose its technological predominance---the backbone of its economic and military power---to China, and should mobilize its intellectuals and allies to shift the tides in its favor \cite{NationalSecurityReport}. The European Union, on the other hand, is planning to spend billions of euros to build a talent pipeline and fund research as part of its \textit{Coordinated Plan on AI} \cite{EU_Report}. Although it is not yet clear who the ultimate leader in the AI arena will be, that leader may well become the world’s next superpower \cite{bostrom2017superintelligence,tegmark2017life}.

This study focuses on scientific collaborations between China-based and U.S.-based scientists in the field of AI. The rationale behind this design is twofold. First, the emergence of China as a leading nation in science \cite{zhou2006emergence,xie2014china} is changing the global balance of power and challenging the U.S.'s decades-long dominance in scientific production \cite{glanzel2008triad,marginson2022all}. In 2020, for example, China produced the largest volume of Science and Engineering publications worldwide (23\% of global output) followed by the U.S.\ (16\%) \cite{NSF2021}. {In general, beyond just AI,} the two countries are leading in terms of the amount spent on research and development (R\&D), e.g., in 2019 the U.S.\ was the world’s leader (\$656 billion) followed by China (\$526 billion), with their expenditures representing 27\% and 22\% of the global total, together representing nearly half of the world's R\&D that year \cite{NSF2022:a}. The gap between these two countries and the rest of the world is even more stark when considering AI venture capital funding in 2020, as 82\% of the year's global investment in this sector went to startups in the U.S.\ (\$27.6 billion) and China (\$16.9 billion) \cite{NSF2022:b}. As we will demonstrate here, Chinese and U.S. AI research reside at the forefront of AI research in terms of novelty and impact.

The second reason is the fierce geopolitical tensions between the U.S.\ and China, which affected scientific collaborations between the two nations. Consider, for example, the policies and investigations launched under the Trump administration, including (i) the China Initiative---a program launched in 2018 by the U.S.\ Department of Justice to counter Chinese national security threats, with a particular emphasis on intellectual property and technology \cite{China:initiative}, and (ii) the investigation of hundreds of scientists by the National Institutes of Health in 2018 \cite{NIH:initiative:a}, with the majority of investigated cases involving receipt of resources from China \cite{NIH:initiative:b}. In a recent study \cite{jia2022impact}, Jia et al.\ found a marked decline in the frequency of U.S.-China collaborations in the life sciences following these investigations. The authors interviewed a number of scientists who had past, ongoing, or planned collaborations with China-based institutions. Those scientists confirmed that, due to the investigations mentioned above, they were reluctant to initiate new or continue existing collaborations with institutions in China. Other recent policies that could potentially limit scientific collaborations between the two nations include the Evaluation of Representative Outcomes (ERO) released by the Chinese government’s Ministry of Science and Technology in 2020 \cite{chinesedocument:2020}, which encourages scholars in China to publish in domestic journals and downplays the importance of international journals \cite{zhong2022china}, hindering cross-border collaborations.

These observations motivate the examination of AI papers produced by the U.S.\ and China, with a particular attention to those resulting from collaboration between the two countries. Here, we analyze a data set of AI papers and scientists to address the following questions: (i) How do the U.S.\ and China rank globally in terms of AI novelty, productivity, and impact? (ii) Do scientists who move from one country to the other continue to collaborate with coauthors from the origin country? If so, at what rate? (iii) Are researchers from the U.S.\ and China more impactful when collaborating together?

\section*{Results}

\subsubsection*{Role of the U.S.\ and China in AI research}

We start our analysis by exploring the countries that lead global AI research. To this end, we utilize Microsoft Academic Graph (MAG) \cite{sinha2015overview}, a widely used dataset containing detailed records of over 263 million scientific publications authored by 271 million scientists. Given our focus on AI research, we focus on papers with the AI field classification according to MAG. For each AI paper, we consider the country of the last author's affiliation as that paper's country of origin. This is based on the convention that the last author is typically ``the head of the lab that hosted most of the research'' \cite{venkatraman2010conventions}. To compare countries in terms of AI research, we focus on four outcome measures: productivity, impact, novelty, and number of scientists. In particular, we measure productivity as the number of AI papers produced by the country. 
In line with common practices in the scientometrics literature, we adopt the approach of AlShebli et al. (2022) \cite{alshebli2022Beijing} to assess impact as the number of citations that AI papers from the country have accumulated within the first two years post-publication. This time frame is widely recognized as an indicator that strongly correlates with the overall citation performance over longer periods.

Next, we examine two types of novelty, which we refer to as context novelty and content novelty. More specifically, context novelty is computed using the z-score measure as proposed by Uzzi et al.~\cite{uzzi2013atypical} while focusing on AI papers. Intuitively, for any given AI paper, the measure considers all pairs of journals (``contexts'') referenced therein, and for each pair, quantifies the likelihood of them being co-cited in AI papers. If this likelihood in our observed data is more than expected by random chance, it indicates relatively common or ``conventional'' pairings. On the other hand, if it is less than expected, it indicates relatively atypical or ``novel'' pairings. As for content novelty, we compute it following Shi and Evans' approach \cite{shi2023surprising}. In particular, we model scientific publications as hypergraphs of keywords (``contents'') and estimate the propensity that a pair of keywords would appear in the same paper. The probability that a set of keywords appears in one publication is calculated as the product of two factors: the proximity between a set of keywords and the popularity of those keywords, which are jointly estimated using a hypergraph embedding algorithm with negative sampling.

Finally, to measure the number of AI scientists who reside in any given country, we first identify AI scientists following the approach of AlShebli et al.\  \cite{alshebli2022Beijing}. In particular, we use the MAG dataset, and classify each publishing researcher as an AI scientist if they satisfy the following conditions: (i) they authored at least three papers; and (ii) the majority of their papers are classified as AI papers. The number of AI scientists in any given country is taken as the number of those whose affiliation resides in that country. Scientists who happen to have simultaneous affiliations in multiple countries are double-counted, i.e., they contribute towards the number of AI scientists in each of these countries. We analyze these outcome measures between the years 2000 to 2021, resulting in a dataset of 5,399,828 papers and 362,929 scientists.

\begin{figure}[H]
\centering
\includegraphics[width=\textwidth]{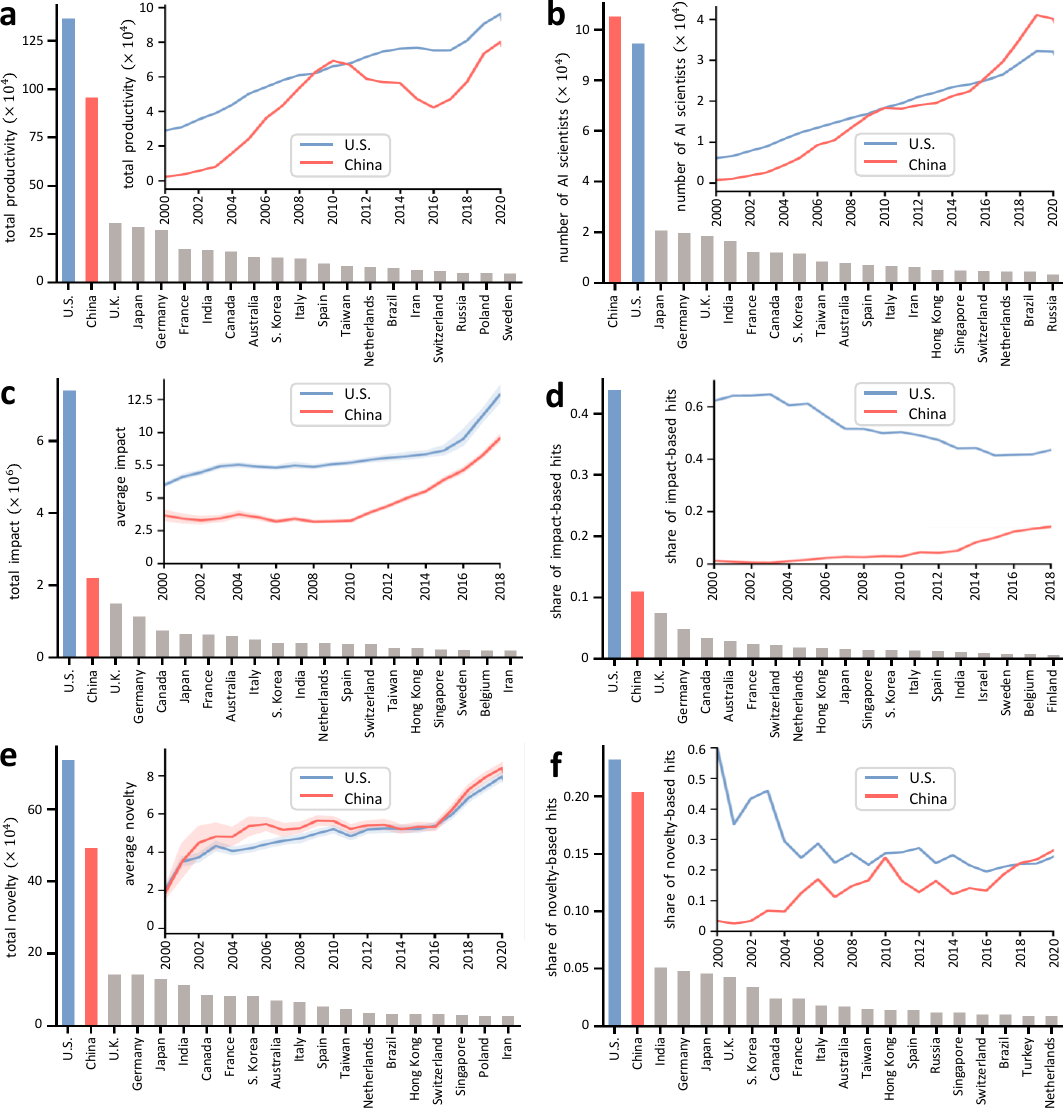}
\caption{\textbf{The U.S.\ and China leading in impact, novelty, productivity, and number of AI scientists.} \textbf{a}, Main plot: The 20 countries producing the most AI papers. Inset: The number of AI papers produced by the leading two countries (U.S.\ and China) over time. \textbf{b}, Main plot: The 20 countries with the largest number of AI scientists. Inset: The number of AI scientists in the leading two countries (U.S.\ and China) over time. \textbf{c}, Main plot: The 20 countries that garner the highest impact in AI research. Inset: The average impact of AI papers in the leading two countries (U.S.\ and China) over time. \textbf{d}, Main plot: The 20 countries with the greatest share of hits based on impact. Inset: The share of hit AI papers based on impact in the leading two countries (U.S.\ and China) over time. \textbf{e} and \textbf{f}, The same as (c) and (d), respectively, but for context novelty instead of impact.}
\label{fig:leaders_in_AI}
\end{figure}

Figure~\ref{fig:leaders_in_AI}a depicts the total AI productivity of the 20 most productive countries. As can be seen, the U.S.\ has produced a total of 1,365,452 AI papers (25.23\% of global output) while China has produced a total of 957,840 papers (17.70\% of global output), demonstrating their global dominance of AI productivity. When looking at the annual productivity of the U.S.\ and China over time, we find that China caught up with the U.S.\ by the year 2010, but has fallen slightly behind in years that followed. Figure~\ref{fig:leaders_in_AI}b focuses on the number of AI scientists in each country, showing that China is leading with 105,103 scientists, compared to the 94,363 in the U.S. This is due to China's substantial growth in AI scientists over the last 5 years, as shown when plotting these numbers over time.

Figure~\ref{fig:leaders_in_AI}c presents the 20 countries with the highest impact in AI research, where impact is quantified based on citations accumulated two years post publication. We find that the U.S.\ and China lead the pack with a total impact of 7,368,464 and 2,157,122 citations, respectively. Taken together, these citations amount to 46.4\% of global impact. Looking at average impact per paper over time, we observe an overall upward trajectory in average impact over time for both countries. To gain deeper insight, we investigate the countries with the largest share of ``hits" based on impact, where a hit is taken as a paper that falls among the 1\% of most impactful papers published in that year \cite{wang2019early}. That is, for each country, we compute the percentage of its papers that are hits out of all AI hits produced globally, rather than from all AI papers produced by that country. Figure~\ref{fig:leaders_in_AI}d illustrates that the U.S.\ has the largest share of AI hits (43.9\% of global hits), followed by China (10.9\%). Looking at each country's share per annum, we find that China's share of hits has increased over time, while the U.S.\ has witnessed a slight decline in its share.

Finally, Figure~\ref{fig:leaders_in_AI}e and Supplementary Figure~1a show that the U.S.\ and China are also at the forefront in terms of context novelty and content novelty, respectively. When examining context novelty over time (Figure~\ref{fig:leaders_in_AI}e), we find that the average novelty of China-based papers has been comparable to, if not greater than, that of U.S.-based papers over the past two decades. In terms of content novelty (Supplementary Figure~1a), we find that it is slightly increasing over time for  both countries, but the gap between the U.S.\ and China persists over time.  Figure~\ref{fig:leaders_in_AI}f and Supplementary Figure~1b also focus on context novelty and content novelty, respectively, but they depict each country's share of novelty hits (i.e., share of papers that fall among the 1\% most novel papers). As can be seen, out of all countries analyzed, the U.S.\ and China have the largest shares. When examining the shares of these two countries over time, we find that China has caught up with the U.S.\ in recent years in terms of context novelty, but remains behind in terms of content novelty.

To determine the degree to which the findings of context novelty are driven by the introduction of relatively new publication venues, we repeated the context novelty analysis while excluding any venues that appear for the first time in the MAG dataset after 2015. Note that the novelty score of any paper published after 2015 may change as a result of this exclusion. We compared the scores of these papers before and after the exclusion of such references; see Supplementary Figure~2. The broad trends remain largely unchanged, suggesting that they cannot be explained by the introduction of newer venues.

\subsubsection*{Cross-border AI scientists' movement and collaboration}

Our exploratory analysis has shown that the U.S.\ and China are leading in AI research. Next, we focus on AI scientists moving from the U.S.\ to China and vice versa.

That is, we analyze individuals who were research active in one of the two countries, before moving to the other. Note that those who were in China, for example, and moved to the U.S.\ before publishing any papers are not considered in our analysis, because they obtained the ``scientist'' status after, not before, the move took place. We rely on affiliations to identify scientists who have moved from one country to another. More specifically, individuals who published AI papers with an affiliation in Country~A, and then started publishing AI papers with an affiliation in Country~B, are considered to have moved from A to B. Scientists who simultaneously held affiliations in multiple countries are excluded from this analysis. Note that our analysis does not study movements from Chinese institutions to American ones, but rather from China to the U.S., as some institutions in China may not necessarily be Chinese.

\begin{figure}[H]
\centering
\includegraphics[width=0.98\textwidth]{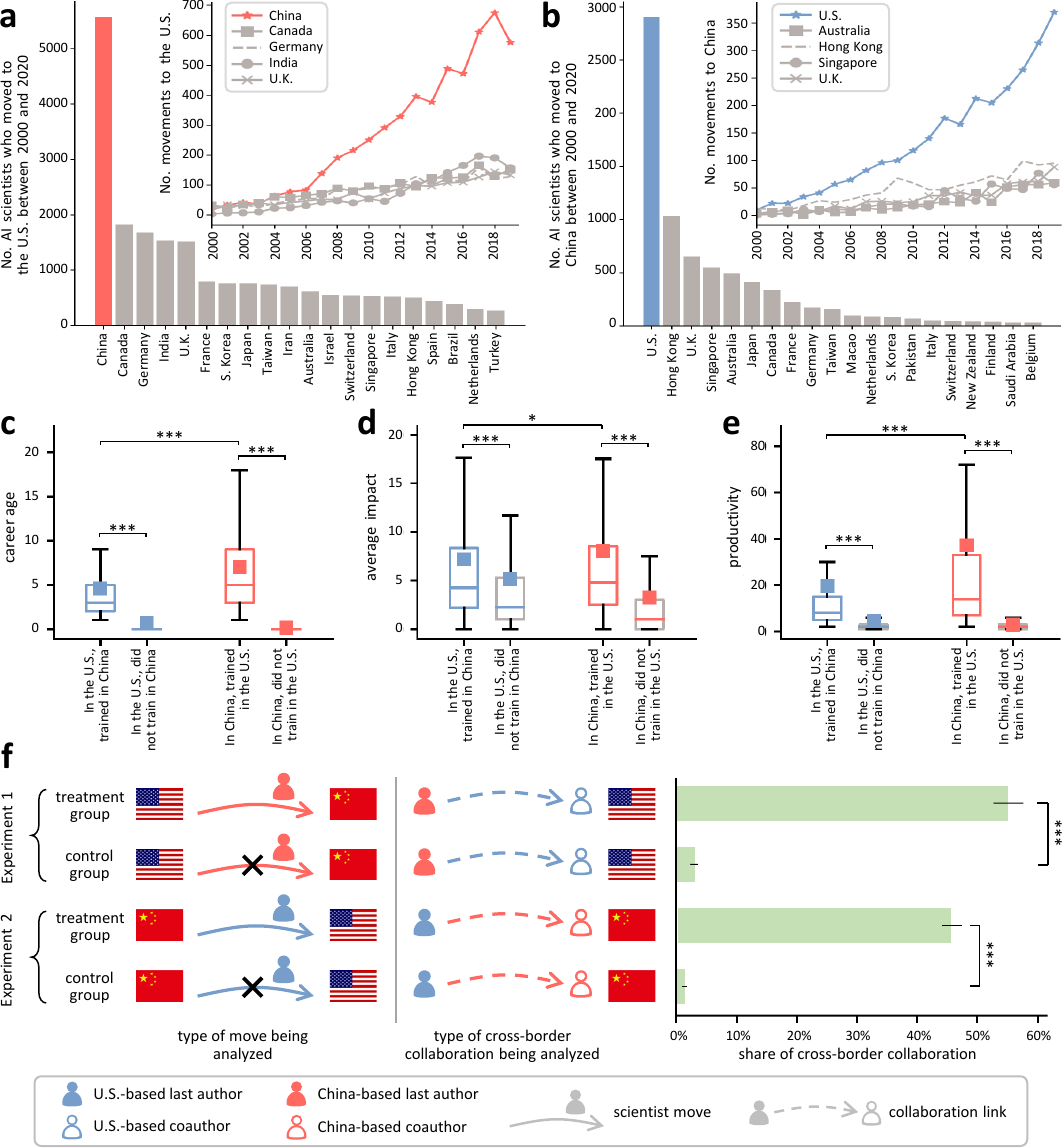}
    \caption{\textbf{The interplay between AI scientists' migration and cross-border collaboration.} \textbf{a}, Main plot: The 20 countries from which the largest number of AI scientists migrate to the U.S. Inset: For the top five countries, the number of AI scientists that migrate to the U.S.\ over time. \textbf{b}, The same as (a) but for China instead of the U.S.
\textbf{c}-\textbf{e}, For AI scientists migrating from the U.S.\ to China, and vice versa, the distributions of career age, impact, and productivity.
\textbf{f}, 
Out of all AI scientists that are based in country $A$, comparing those who migrated from country $B$ to those who did not, in terms of the percentage of their papers that involve coauthors from $B$; the comparison is done using Coarsened Exact Matching, and only considers the years after the migration, while controlling for career age, impact, and productivity. Error bars represent bootstrapped 95\% confidence intervals; P values are from t-tests; * $p<.05$; *** $p<.001$.
} 
\label{fig:migration}
\end{figure}

Results from this analysis are summarized in Figure~\ref{fig:migration}. In particular, Figure~\ref{fig:migration}a examines the distribution of countries from which AI scientists moved to the U.S.\ during the past 20 years, showing that most of them come from China. Moreover, the number of AI scientists moving from China to the U.S.\ has increased rapidly over the past decade. Similarly, as shown in Figure~\ref{fig:migration}b, the country from which AI scientists move the most to China is the U.S., and the number of such movements has increased steadily over the past two decades. Figures~\ref{fig:migration}c to \ref{fig:migration}e show that China is attracting AI scientists who are more experienced, impactful, and productive than those attracted to the U.S. These figures also show that China-based scientists who move from the U.S. have greater experience, impact, and productivity than those who do not. The same holds for U.S.-based scientists who move from China compared to those that do not.

To better understand the nature of such movements, we infer the race of the scientists moving from China to the U.S. and vice versa, using a name-ethnicity classifier. More specifically, we use \textit{NamePrism}, which classifies scientists into six different racial groups: Asian / Pacific Islander (API), American Indian / Alaskan Native (AIAN), Black, Hispanic, Two or more races (2PRACE), and White~\cite{ye2017nationality}. Note that NamePrism is widely used in the social sciences to infer the race or ethnicity of given names~\cite{alshebli2018preeminence, ghosh2021fair, zeina2020gender, o2022ethnic, law2022public,liu2023non}. We find that 98.8\% of scientists moving from China to the U.S., and 96.0\% of those moving from the U.S. to China have Asian ancestry. This suggests that the predominant flow of movement between these two nations is among individuals of Asian descent.

{Next, we examine the relationship between the university ranking and the rate at which scientists move to, or from, that university. We find that 27\% of AI scientists moving from the U.S.\ to China come from a top-100 institution. Similarly, 29\% of those moving in the opposite direction come from a top-100 institution. As such, the two nations seem comparable in the rate at which they attract AI scientists from the other nation's top institutions. However, the nations differ considerably in the rate at which their top institutions recruit talent from the other nation. In particular, 37\% of those who move from China to the U.S.\ are recruited by a top-100 institution, whereas the rate is only 20\% of those moving in the opposite direction.}

Finally, we compare China-based AI scientists who move from the U.S.\ to those who do not, in terms of the rate at which they collaborate with U.S.-based coauthors after the movement took place. Likewise, we compare U.S.-based scientists who move from China to those who do not, in terms of the rate at which they collaborate with China-based coauthors. The comparison is carried out using Coarsened Exact Matching while controlling for (i) career age, (ii) productivity in the movement year, and (iii) citations accumulated by the movement year. Each matching experiment compares the treatment group (scientists in country $A$ who moved from country $B$) to the control group (scientists in $A$ who did not move from $B$) in terms of the rate at which they collaborate with coauthors from $B$ in the years following the movement. {Naturally, one would expect the rate to be higher for the treatment group. Nevertheless, this analysis allows us to quantify the difference in the rate between the two groups.}

The two matching experiments and their outcomes are illustrated in Figure~\ref{fig:migration}f; see Supplementary Table~1 for numeric values. As shown in this figure, China-based AI scientists who moved from the U.S.\ are nearly 20 times more likely to collaborate with U.S.-based coauthors, compared to their counterparts who did not move from the U.S.\ (54.45\% vs.\ 2.78\%). Similarly, the likelihood of U.S.-based AI scientists who moved from China to collaborate with China-based coauthors is nearly 30 times greater, compared to their counterparts who did not move from China (42.21\% vs.\ 1.46\%). {Furthermore, we examined the relationship between the academic age of the AI scientist and the rate at which they collaborate with the country of origin (i.e., the country that they have moved from). We found a weak correlation ($r = 0.22$; $p < 0.001$) between academic age and collaboration rate, but only when the move is from China to the U.S. (as for the other direction, no significant correlation was found).}

{Naturally, one would expect the ``effect'' of the relocation to be strongest in the time immediately after the relocation, and fade thereafter as local collaborations put out international ones with costly communication (e.g., time zone differences between the U.S.\ and China). To determine whether this is the case, we examined the rate at which scientists collaborate with the country of origin over time, focusing on the 10 years that followed the move. Indeed, as shown in Supplementary Figure~3, the effect fades out for China-based scientists, dropping from 57\% (immediately after they relocate to China) to about 36\% a decade after the move. Interestingly, however, the opposite is true for U.S-based scientists; the rate at which they collaborate with China-based coauthors actually increases (from 45\% to 51\%) in the decade that follows their move to the U.S.}

\subsubsection*{Collaborative AI research between the U.S. and China}

Our final analysis focuses on papers produced when the two countries collaborate and compares them to those produced when each country works without the other. {International collaborations have been shown to provide rich opportunities to enhance research \cite{jeong2014drivers}, as they free scientists from local constraints \cite{wagner2005network} and offer diverse views and experiences, thereby providing a competitive advantage \cite{francisco2015international}. However, the rivalry between the U.S.\ and China, especially in the field of AI, raises several questions about the international collaborations between the two nations in this field. For instance, is the number of such collaborations decreasing in recent years due to the aforementioned rivalry? Are such collaborations taking place mainly among higher-ranked universities? And if there is indeed a citation premium for such collaborations, is it greater in the field of AI compared to other fields? and can it be explained by ``home citations''?}

{To explore these questions}, for any given paper, if the last author is affiliated with an institution from the U.S., and at least one coauthor is affiliated with an institution from China, we consider this to be a U.S.-based paper produced in collaboration with China. Similarly, if the last author has a China-based affiliation, and at least one coauthor has a U.S.-based affiliation, we consider this to be a China-based paper produced in collaboration with the U.S. This approach is similar to the one used by AlShebli et al.\ \cite{alshebli2022Beijing}, except that they focus on cities, while our focus is on countries. Given our reliance on the order of authorship to infer lead author, it is not possible to detect cases where there are multiple leads. Moreover, since we focus on papers either led by China or led by the U.S., any paper involving a China-based coauthor and a U.S.-based coauthor is only considered a collaboration between the two countries if one of the two coauthors is a last author.

\begin{figure}[H]
\centering
\includegraphics[width=\textwidth]{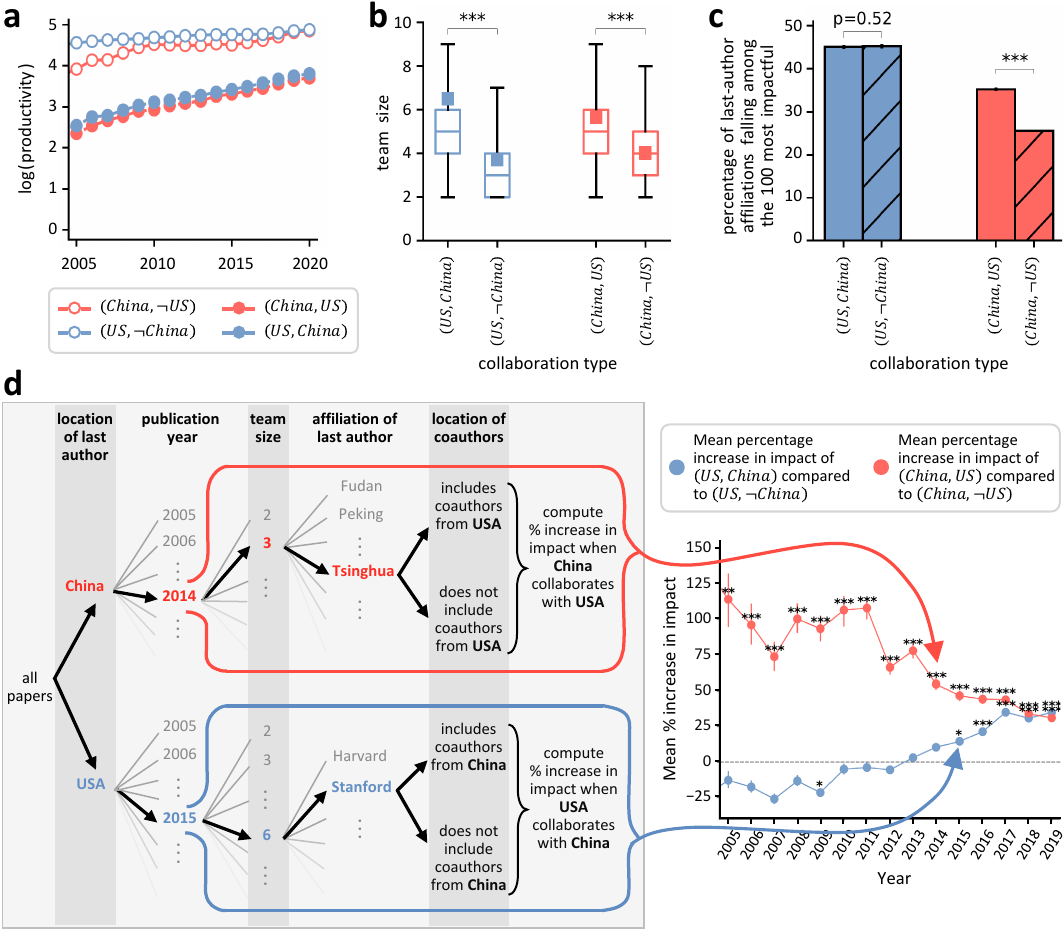}
\caption{\textbf{The impact of U.S.-China AI collaborations.}
Analyzing four types of AI papers: (i) U.S.-based papers produced in collaboration with China, $(\mathit{US}, \mathit{China})$; (ii) U.S.-based papers produced without China, $(\mathit{US}, \neg\mathit{China})$; (iii) China-based papers produced in collaboration with the U.S., $(\mathit{China}, \mathit{US})$; (iv) China-based papers produced without the U.S., $(\mathit{China}, \neg\mathit{US})$.
\textbf{a}, Annual number of papers per type (log scale).
\textbf{b}, Distribution of team size per type.
\textbf{c}, For each type, the percentage of papers of which the last author's affiliation is among the 100 most impactful institutions in AI. 
\textbf{d}, The infographic illustrates the Coarsened Exact Matching (CEM) experiment, while the plot depicts the percentage increase in impact of $(\mathit{US}, \mathit{China})$ compared to $(\mathit{US}, \neg\mathit{China})$, as well as the percentage increase in impact of $(\mathit{China}, \mathit{US})$ compared to $(\mathit{China}, \neg\mathit{US})$, over time.
Error bars represent bootstrapped 95\% confidence intervals; P values are calculated using t-tests (a, b, and d) and two-sided Fisher’s exact test (c); * $p<.05$; ** $p<.01$; *** $p<.001$.
}
\label{fig:china_us_collab}
\end{figure}

To be more precise, we compare U.S.-based papers produced in collaboration with China, denoted by $(\mathit{US}, \mathit{China})$, to those produced without such collaboration, denoted by $(\mathit{US}, \neg\mathit{China})$. We also compare China-based papers produced in collaboration with the U.S., $(\mathit{China}, \mathit{US})$, to those produced without such collaboration, $(\mathit{China}, \neg\mathit{US})$. More specifically, Figure~\ref{fig:china_us_collab}a depicts the number of such papers over time on a log scale. As can be seen, collaborations between the U.S.\ and China in the field of AI were relatively rare before 2010 (fewer than 1000 papers per year), and started increasing in the second decade of the millennium. Despite this increase, collaborations between the two countries continue to represent only a small fraction of their overall AI productivity. 

Figure~\ref{fig:china_us_collab}b compares the number of authors on these papers. As shown in this figure, teams that involve collaborations between the two countries are, on average, larger than those that do not.
Figure~\ref{fig:china_us_collab}c compares these papers in terms of the percentage of last-author affiliations that fall among the top 100 most impactful institutions in the field of AI. As can be seen, last authors of China-based papers are more likely to be affiliated with a top-100 institution when collaborating with the U.S.

Figures~\ref{fig:china_us_collab}a to \ref{fig:china_us_collab}c imply that, when comparing the impact of papers that involve a collaboration between the U.S.\ and China to those that do not, one needs to control for publication year, team size, and last-author affiliation. To this end, we use Coarsened Exact Matching (CEM) \cite{iacus2012causal}. The matching process is illustrated in the left-hand side of Figure~\ref{fig:china_us_collab}d, emphasizing that the impact of papers in $(\mathit{US}, \mathit{China})$ is compared to the impact of those in $(\mathit{US}, \neg\mathit{China})$ that have the same publication year, team size, and last-author affiliation. Likewise, the impact of papers in $(\mathit{China}, \mathit{US})$ is compared to the impact of those in $(\mathit{China}, \neg\mathit{US})$ while controlling for the above confounders. The right-hand side of Figure~\ref{fig:china_us_collab}d depicts the relative difference in impact between the papers and their matched ones over time; see Supplementary Tables 2 and 3 for the numeric values. As can be seen, for the entirety of the period considered in our analysis, China-based papers that involve U.S.-based collaborators have been more impactful than those that do not. As for U.S.-based papers, the first decade in our analysis shows no difference in impact when involving China-based scientists. However, the last five years have witnessed a significant increase in impact associated with collaborators from China. Similarly, when examining the rate at which the two countries produce impact-based AI hits, we find that each of them achieves a notably higher rate when collaborating with the other; see Supplementary Figure~4.

{As a robustness check, we perform a within-subject analysis by controlling for the last author. That is, we compare papers that involve U.S.-China collaborations to those produced by one country without the other, but the comparison is now performed among papers that have the same last-author. This naturally yields a smaller dataset compared to our original one, since there are many scientists who do not meet this requirement. To account for the smaller dataset, we relaxed our matching criteria by allowing for up to two years difference in publication date, and binning the sizes of teams that involve five or more members. The results of this analysis are depicted in Supplementary Figure~5. As can be seen, our main findings persist, i.e., both countries yield more impactful AI research when collaborating with each other, especially in recent years.}

{Our next robustness check involves expanding the definition of ``AI papers'' by including those classified as Computer Vision, Pattern Recognition, Machine Learning, or Natural Language Processing (NLP). Due to the relatively small number of NLP papers in the MAG dataset that include U.S.-China collaborations, we opted to merge NLP with Machine Learning. For each of the resulting subfields, we replicated Figure~\ref{fig:china_us_collab}d, examining the percentage increase in impact when the U.S.\ and China collaborate together. Additionally, we replicated the figure, but this time combining all subfields together with all the papers from our original dataset. The results of these analyses confirm our main finding, i.e., the two nations are more impactful when collaborating together; see Supplementary Figure~6.}

We performed yet another robustness check, where we explore an alternative way of identifying collaborations between the two countries. Instead of the lead author being in one country and a coauthor being in the other (regardless of the coauthor's position on the paper), we now require the coauthor to be the first author on the paper. This yields a stricter criterion than the one used earlier, as it effectively excludes all collaborations where the coauthor is not the first author. This suggests that a scientist from one country has ``done the most work'', and a scientist from the other country has ``provided the most guidance''. The result of this analysis is presented in Supplementary Figure~7. As can be seen, the same broad trends are observed, i.e., China and the U.S. tend to produce more impactful AI research when collaborating together.

Our citation-based analysis focused on two years post publication in order to include recent cohorts of papers. Including such papers is particularly relevant in our context, because AI is a fast-moving field, and the number of AI papers generated in both the U.S. and China have rapidly increased over time. Thus, by limiting the number of years analyzed to just two, we can include the most recent developments in the field. There have been debate regarding the use of short-term impact as a predictor of long-term impact, however; see, e.g., \cite{wang2014comment}. This controversy considered only papers published in \textit{Physical Review}; the extent to which this holds in the field of AI remains unknown. Against this background, we computed the correlation between the number of citations received within two years, and those received within five years; we found a strong correlation between the two (Pearson’s $r$ = 0.85). Additionally, as a robustness check, we repeated the same analysis of Figure~\ref{fig:china_us_collab}d, while focusing on five years instead of two; see Supplementary Figure~8. Again, the same broad trends are observed, i.e., papers from China that involve U.S.-based collaborators receive more citations than those that do not. Similarly, in recent years, papers from the U.S. that involve China-based collaborators receive more citations than those that do not.

To determine whether the observed trend (i.e., the citation premium of U.S.-China collaborations in recent years) is specific to AI, or is a reflection of a general U.S.-China collaboration, we performed the same Coarsened Exact Matching analysis, but for seven additional fields of Computer Science, namely Algorithms, Computer Networks, Computer Security, Databases, Operating Systems, Programming Languages, and World Wide Web. The results of this analysis are depicted in Supplementary Figure~9. As can be seen in this figure, the citation premium associated with U.S.-China collaborations is greater in AI than any of the other Computer Science fields considered in our analysis. This finding suggests that such collaborations matter more in AI than in other fields.

One possible explanation behind these trends could be the prevalence of ``home citations'', i.e., the citations received from papers produced by the same country. If scholars consider how citations are used to rank scholars and institutions, and if their access to government resources depends on raising their international profiles, then they may have an incentive to ``game the system'' by deliberately adopting practices that raise their measured impact. In particular, scholars may follow a conscious and deliberate practice of citing one another in an effort to improve their measured academic output and that of their home country institutions. To explore this possibility, we repeat the same analysis while focusing solely on the citations made by non-U.S.\ papers to U.S.-based work, as well as those made by non-China papers to China-based work. In this context, we use two alternative ways of identifying the non-U.S.\ and non-China papers. The first focuses on the last author (e.g., the paper is non-U.S.\ if the last author does not have a U.S.-based affiliation), while the second considers all authors (e.g., the paper is non-U.S.\ if none of the affiliations therein are U.S.-based). The results of this analysis are depicted in Supplementary Figure~10. As shown in this figure, the same broad trends are observed, suggesting that they cannot be entirely explained by home citations.

\begin{figure}[H]
\centering
\includegraphics[width=\textwidth]{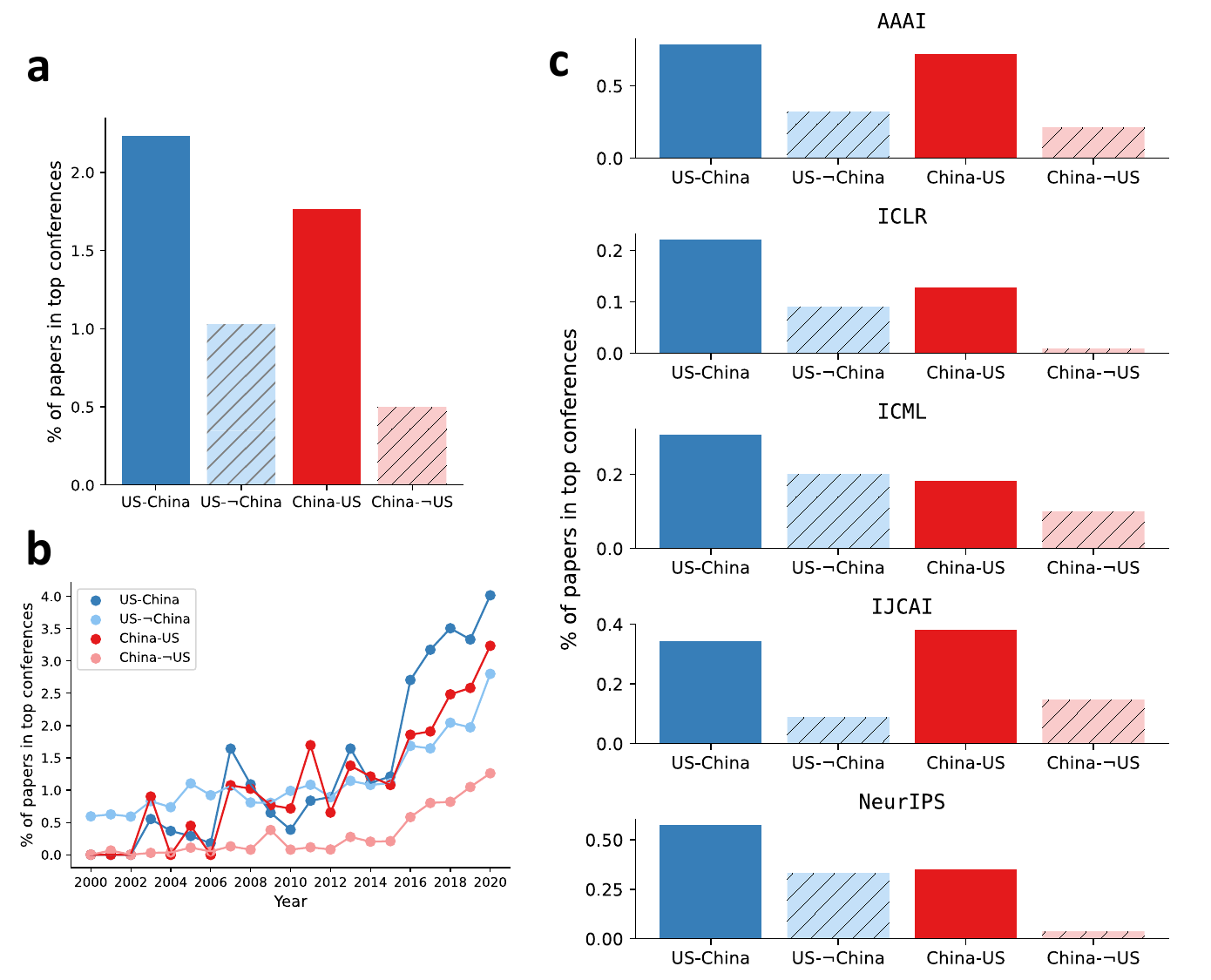}
\caption{\textbf{Publication venue analysis.} Comparing four types of AI papers: (i) U.S.-based papers produced in collaboration with China, $(\mathit{US}, \mathit{China})$; (ii) U.S.-based papers produced without China, $(\mathit{US}, \neg\mathit{China})$; (iii) China-based papers produced in collaboration with the U.S., $(\mathit{China}, \mathit{US})$; (iv) China-based papers produced without the U.S., $(\mathit{China}, \neg\mathit{US})$. \textbf{a}, For each type, the percentage of papers published in the top five AI conferences. \textbf{b}, Similar to (a) but over time. \textbf{c}, Similar to (a) but disaggregated across the top five AI conferences.
}
\label{fig:china_us_venues}
\end{figure}

Up to this point in our analysis of U.S.-China collaborations, we focused on citation count. Next, we focus on an alternative measure---the publication rate in top AI conferences. To this end, we use CORE, a website providing conference rankings that are widely used by computer scientists \cite{CORE}. According to CORE, ``conference rankings are determined by a mix of indicators, including citation rates, paper submission and acceptance rates, and the visibility and research track record of the key people hosting the conference and managing its technical program''. The highest rank is ``A*'', and only 7\% of computer science conferences receive this highly prestigious ranking \cite{CORE}. Importantly, these conferences are double-blind, meaning that reviewers cannot see the authors' names or affiliations, unlike journal submissions. Since CORE does not specify which of the A* conferences is AI-related, we identified the 10 most highly ranked AI venues according to Google Metrics, and found five that are A* conferences (the other five venues were journals); these conferences are: AAAI, ICLR, ICML, IJCAI, and NeurIPS. When analyzing the papers published in these conferences, we focus on the same four types of papers used earlier, i.e., $(\mathit{US}, \mathit{China})$, $(\mathit{US}, \neg\mathit{China})$, $(\mathit{China}, \mathit{US})$, and $(\mathit{China}, \neg\mathit{US})$. For each type, we compute the percentage of papers produced by that type that get published in each of the top five conferences; see Figure~\ref{fig:china_us_venues}a. As shown in this figure, $(\mathit{US}, \mathit{China})$ and $(\mathit{China}, \mathit{US})$ publish in the top five A* conferences at a relatively higher rate than $(\mathit{US}, \neg\mathit{China})$ and $(\mathit{China}, \neg\mathit{US})$, respectively. When considering these rates over time, we find that the publication premium of U.S.-China collaborations has become particularly pronounced in recent years (Figure~\ref{fig:china_us_venues}b). Finally, when considering the conferences in isolation, we find that the publication premium is greater in each of the five Computer Science conferences that are considered in our analysis  (Figure~\ref{fig:china_us_venues}c). These results further support our finding that China and the U.S.\ produce more impactful AI research when collaborating together.

\section*{Discussion}

Our study contributes to a growing body of work that focuses on AI bibliometrics. For example, Frank et al.~\cite{frank2019evolution} examined research that interacts with AI using the MAG dataset to identify disciplines frequently citing or cited by AI papers. Similarly, Tran et al.~\cite{tran2019global} examined AI dynamics in medicine, analyzing diseases most frequently studied most and least in AI using the Web of Science. Martínez-Plumed et al.~\cite{martinez2021research} identified research communities most likely to advance the state of the art in AI using Papers With Code---a repository of AI benchmarks along with their associated papers. In related research, Tang et al.~\cite{tang2020pace} gained insight into how the pace of AI innovations has changed over the past years by analyzing preprints on arXiv.org. Klinger et al.~\cite{klinger2021deep} identified the AI subfields most involved in the development of Explainable AI---an area of research that focuses on making AI-based decisions more understandable to humans using bibliometric data from Scopus. Finally, Stathoulopoulos and Mateos-Garcia probed gender differences in the authorship of AI papers \cite{stathoulopoulos2019gender}. Despite the number and variety of these analyses, U.S.-China collaborations in the field of AI have not been closely examined to date. Our study fills this gap in the literature, by examining the migration of AI scientists from one country to the other, and by analyzing the impact of AI papers resulting from collaborations between the two countries.

Our study also contributes to another line of research that analyzes the mobility of scientists. In particular, some scholars have modeled this phenomenon \cite{deville2014career,james2018prediction,vaccario2021reproducing} and identified factors that influence scientists' migration decisions \cite{van2012science,franzoni2015international,appelt2015factors,azoulay2017mobility}, while others have focused on the impact of migration on scientists' careers \cite{jonkers2013research,scellato2015migrant,baker2015non,petersen2018multiscale} or countries' research activities \cite{doria2017quantifying,wagner2018openness,chinchilla2018global,verginer2020cities}. Some of this work has focused on the mobility of scientists from specific disciplines, e.g., physics \cite{deville2014career,dyachenko2017internal,petersen2018multiscale}, economics \cite{baker2015non}, genomics \cite{huang2014mobility}, chemical engineering \cite{slavova2016learning}, life sciences \cite{azoulay2017mobility}, stem cell research \cite{levine2006research}, biology, chemistry, material science, and environmental science \cite{scellato2015migrant}. The mobility of AI scientists has only been considered very recently by AlShebli et al.~\cite{alshebli2022Beijing}, who constructed networks of inter-city AI citations, collaborations, and migrations. Using these networks, the authors quantified the role that different cities play in bridging East and West, and showed that Beijing's role surpasses that of all other cities combined, becoming a global hub for AI researchers. Moreover, the authors tracked the AI center of mass---the average geographic location of AI research activities---over the past three decades, and found that it has been drifting steadily towards Asia, with Beijing's gravitational pull increasing each year, suggesting an eastward shift in the tides of AI research. Our study complements this literature, showing that the majority of AI scientists moving to the U.S.\ come from China, and vice versa. Moreover, those who move continue to collaborate with scientists in the origin country. Our findings imply that the movement of AI scientists between the U.S.\ and China fosters fruitful collaborations that can be advantageous to both countries.

Studies examining international collaborations in general (rather than in the field of AI) suggest that the impact of international collaborations varies across countries \cite{glanzel2001national}. Such collaborations are associated with greater impact for UK-based scientists \cite{katz1997much} as well as Norway-based scientists \cite{aksnes2003characteristics}. Conversely, most highly cited papers from the U.S.\ are produced by domestic, rather than international, collaborations \cite{persson2010highly}. Similarly, Kwiek \cite{kwiek2021large} analyzed papers from six fields of research---engineering and technologies, agricultural sciences, humanities, natural sciences, medical sciences and social sciences---and found that, for both the U.S.\ and China, domestic collaborations (i.e., those involving U.S.-based only, or China-base only, coauthors) have a greater impact than their international counterparts. In contrast, our study shows that when it comes to AI research, papers involving collaborations between the U.S.\ and China (whether U.S.- or China-led) are more impactful than those led by one of the two countries and not involving the other. 

We focused on the U.S. and China in our analyses given the geopolitical tension between the two, and that they are the most productive countries in terms of AI research. Nevertheless, the synergy we found between the U.S. and China is not unique; similar synergies also exist when these nations collaborate with other countries. We examined this possibility by examining their collaborations with the most productive countries in AI research. Results from this analysis are depicted in Supplementary Figure~11. This reveals that the U.S. produces more impactful AI research when collaborating with Australia, the UK, and Canada. Similarly, China produces more impactful AI research when collaborating with India, South Korea, Australia, the UK, Canada, and France. The most impactful collaborations with the U.S. are with China, but the converse is different. In particular, China's collaborations with the U.S. are less impactful than those with India, South Korea, Australia, the UK, and Canada. This exploratory analysis raises the possibility that the U.S. may have more to lose than China from policies discouraging collaborations between the two nations, as China would continue to have more powerful collaborators, whereas the U.S. would lose its most powerful one. Further research, however, is needed to examine this possibility in greater depth.

{Our study is not without limitations. First, although we rely on a widely used dataset (namely Microsoft Academic graph), this dataset is historically biased toward English-speaking cultures, and lacks papers written in Mandarin or Cantonese, thereby underrepresenting the research output of China. Second, since our dataset ends in 2021, it does not capture the latest developments in the field of AI such as, e.g., the recent rise of large language models (LLMs), which may have dramatically shifted  the fields of AI and NLP. As such, it remains to be seen whether the U.S.-China race to innovate in the space of LLMs exhibits similar trends to the ones reported in our study. Third, when quantifying impact, we only consider citations that the paper has accumulated in the two years (Figure~\ref{fig:china_us_collab}d, as per \cite{alshebli2022Beijing}) or five years (Supplementary Figure~8, as per \cite{alshebli2018preeminence}) post publication. As such, our analysis does not highlight ``sleeping beauties'' \cite{van2004sleeping} that start accumulating papers at a later stage (e.g., a decade after publication, as per \cite{sinatra2016quantifying}). Fourth, when comparing the U.S.\ and China in terms of the number of AI scientists, the comparison is done in absolute rather than relative terms, i.e., we do not account for the substantial difference in population size between the two nations. Fifth, our approach of inferring scientists' Asian ancestry cannot capture the qualitative nuance of scientists migrating from the U.S.\ to China or vise versa, e.g., those who do not wish to identify with their prior ancestry.}.

Future extensions could look beyond English-centric data sources, by examining bibliometric datasets that include papers written in Mandarin or Cantonese \cite{CSCD,CSSCI}, thereby providing a more comprehensive view of U.S.-China research collaboration. Additionally, future research could examine the influence of broader social dynamics and terminological variations on migration trends and research outputs, focusing on the history of AI winters and how it influenced both the migration of AI researchers as well as the rebranding of AI terminology to fit different funding schemes \cite{galanos2022nomadic}. Future work may also enrich our analysis by contextualizing the long-standing history of AI development and collaboration between China and the U.S., including the military and defense implications of AI, which underscore the dual-use nature of such technologies, complicating the narrative of academic collaboration \cite{galanos2023expectations}.

{Analysis of AI's potential impact on U.S.-China relations is heavily influenced by technonationalism, which emphasizes interstate competition over technological assets \cite{ding2022dueling}. For instance,} the National Security Commission on AI warns that the U.S.\ ``\textit{must win the AI competition that is intensifying strategic competition with China}'' and that ``\textit{China’s plans, resources, and progress should concern all Americans}'' \cite{NationalSecurityReport}. Such portrayal of China as a threat to the U.S.\ in the AI arena may hinder scientific collaborations between the two nations. Indeed, these collaborations can be viewed from a national perspective, where the emphasis is on increasing economic competitiveness, ensuring national security, and signaling national prestige. After all, the line between basic and commercial research is thinning, and this is challenging conventional norms about international collaboration and the nature of scientific ownership \cite{daniels2022knowledge}. Nevertheless, despite potential spillovers that might work against national interests, such collaborations can also be viewed from a perspective that transcends political agendas---a global perspective where the emphasis is on the advancement of knowledge for the benefit of all.

\section*{Data and Code Availability}

All data generated or analysed in this study can be downloaded from the Microsoft Academic Graph (MAG) {\href{https://www.microsoft.com/en-us/research/project/microsoft-academic-graph/}{website}}. All code used to produce the figures and analyses can be downloaded {\href{https://github.com/samemon/US_China_AI_collaboration}{here}}. 

\bibliography{main.bib}

\begin{thebibliography}{100}
\expandafter\ifx\csname url\endcsname\relax
  \def\url#1{\texttt{#1}}\fi
\expandafter\ifx\csname urlprefix\endcsname\relax\def\urlprefix{URL }\fi
\providecommand{\bibinfo}[2]{#2}
\providecommand{\eprint}[2][]{\url{#2}}

\bibitem{frey2017future}
\bibinfo{author}{Frey, C.~B.} \& \bibinfo{author}{Osborne, M.~A.}
\newblock \bibinfo{title}{The future of employment: How susceptible are jobs to computerisation?}
\newblock \emph{\bibinfo{journal}{Technological forecasting and social change}} \textbf{\bibinfo{volume}{114}}, \bibinfo{pages}{254--280} (\bibinfo{year}{2017}).

\bibitem{frank2018small}
\bibinfo{author}{Frank, M.~R.}, \bibinfo{author}{Sun, L.}, \bibinfo{author}{Cebrian, M.}, \bibinfo{author}{Youn, H.} \& \bibinfo{author}{Rahwan, I.}
\newblock \bibinfo{title}{Small cities face greater impact from automation}.
\newblock \emph{\bibinfo{journal}{Journal of the Royal Society Interface}} \textbf{\bibinfo{volume}{15}}, \bibinfo{pages}{20170946} (\bibinfo{year}{2018}).

\bibitem{acemoglu2018race}
\bibinfo{author}{Acemoglu, D.} \& \bibinfo{author}{Restrepo, P.}
\newblock \bibinfo{title}{The race between man and machine: Implications of technology for growth, factor shares, and employment}.
\newblock \emph{\bibinfo{journal}{American Economic Review}} \textbf{\bibinfo{volume}{108}}, \bibinfo{pages}{1488--1542} (\bibinfo{year}{2018}).

\bibitem{frank2019toward}
\bibinfo{author}{Frank, M.~R.} \emph{et~al.}
\newblock \bibinfo{title}{Toward understanding the impact of artificial intelligence on labor}.
\newblock \emph{\bibinfo{journal}{Proceedings of the National Academy of Sciences}} \textbf{\bibinfo{volume}{116}}, \bibinfo{pages}{6531--6539} (\bibinfo{year}{2019}).

\bibitem{felten2021occupational}
\bibinfo{author}{Felten, E.}, \bibinfo{author}{Raj, M.} \& \bibinfo{author}{Seamans, R.}
\newblock \bibinfo{title}{Occupational, industry, and geographic exposure to artificial intelligence: A novel dataset and its potential uses}.
\newblock \emph{\bibinfo{journal}{Strategic Management Journal}}  (\bibinfo{year}{2021}).

\bibitem{chen2021automation}
\bibinfo{author}{Chen, H.~C.} \emph{et~al.}
\newblock \bibinfo{title}{Automation impacts on china's polarized job market}.
\newblock \emph{\bibinfo{journal}{Journal of Computational Social Science}} \bibinfo{pages}{1--19} (\bibinfo{year}{2021}).

\bibitem{luengo2020artificial}
\bibinfo{author}{Luengo-Oroz, M.} \emph{et~al.}
\newblock \bibinfo{title}{Artificial intelligence cooperation to support the global response to covid-19}.
\newblock \emph{\bibinfo{journal}{Nature Machine Intelligence}} \textbf{\bibinfo{volume}{2}}, \bibinfo{pages}{295--297} (\bibinfo{year}{2020}).

\bibitem{kourou2015machine}
\bibinfo{author}{Kourou, K.}, \bibinfo{author}{Exarchos, T.~P.}, \bibinfo{author}{Exarchos, K.~P.}, \bibinfo{author}{Karamouzis, M.~V.} \& \bibinfo{author}{Fotiadis, D.~I.}
\newblock \bibinfo{title}{Machine learning applications in cancer prognosis and prediction}.
\newblock \emph{\bibinfo{journal}{Computational and structural biotechnology journal}} \textbf{\bibinfo{volume}{13}}, \bibinfo{pages}{8--17} (\bibinfo{year}{2015}).

\bibitem{hosny2018artificial}
\bibinfo{author}{Hosny, A.}, \bibinfo{author}{Parmar, C.}, \bibinfo{author}{Quackenbush, J.}, \bibinfo{author}{Schwartz, L.~H.} \& \bibinfo{author}{Aerts, H.~J.}
\newblock \bibinfo{title}{Artificial intelligence in radiology}.
\newblock \emph{\bibinfo{journal}{Nature Reviews Cancer}} \textbf{\bibinfo{volume}{18}}, \bibinfo{pages}{500--510} (\bibinfo{year}{2018}).

\bibitem{fleming2018artificial}
\bibinfo{author}{Fleming, N.}
\newblock \bibinfo{title}{How artificial intelligence is changing drug discovery}.
\newblock \emph{\bibinfo{journal}{Nature}} \textbf{\bibinfo{volume}{557}}, \bibinfo{pages}{S55--S55} (\bibinfo{year}{2018}).

\bibitem{van2006impact}
\bibinfo{author}{Van~Arem, B.}, \bibinfo{author}{Van~Driel, C.~J.} \& \bibinfo{author}{Visser, R.}
\newblock \bibinfo{title}{The impact of cooperative adaptive cruise control on traffic-flow characteristics}.
\newblock \emph{\bibinfo{journal}{IEEE Transactions on intelligent transportation systems}} \textbf{\bibinfo{volume}{7}}, \bibinfo{pages}{429--436} (\bibinfo{year}{2006}).

\bibitem{spieser2014toward}
\bibinfo{author}{Spieser, K.} \emph{et~al.}
\newblock \bibinfo{title}{Toward a systematic approach to the design and evaluation of automated mobility-on-demand systems: A case study in {S}ingapore}.
\newblock In \emph{\bibinfo{booktitle}{Road vehicle automation}}, \bibinfo{pages}{229--245} (\bibinfo{publisher}{Springer}, \bibinfo{year}{2014}).

\bibitem{ecola2018road}
\bibinfo{author}{Ecola, L.}, \bibinfo{author}{Popper, S.~W.}, \bibinfo{author}{Silberglitt, R.} \& \bibinfo{author}{Fraade-Blanar, L.}
\newblock \bibinfo{title}{The road to zero: A vision for achieving zero roadway deaths by 2050}.
\newblock \emph{\bibinfo{journal}{Rand health quarterly}} \textbf{\bibinfo{volume}{8}} (\bibinfo{year}{2018}).

\bibitem{erel2018could}
\bibinfo{author}{Erel, I.}, \bibinfo{author}{Stern, L.}, \bibinfo{author}{Chenhao, T.} \& \bibinfo{author}{Weisbacj, M.}
\newblock \bibinfo{title}{Could machine learning help companies select better board directors?}
\newblock \emph{\bibinfo{journal}{Harvard Business Review}}  (\bibinfo{year}{2018}).

\bibitem{mullainathan2019biased}
\bibinfo{author}{Mullainathan, S.}
\newblock \bibinfo{title}{Biased algorithms are easier to fix than biased people}.
\newblock \emph{\bibinfo{journal}{The New York Times}}  (\bibinfo{year}{2019}).

\bibitem{kleinberg2018human}
\bibinfo{author}{Kleinberg, J.}, \bibinfo{author}{Lakkaraju, H.}, \bibinfo{author}{Leskovec, J.}, \bibinfo{author}{Ludwig, J.} \& \bibinfo{author}{Mullainathan, S.}
\newblock \bibinfo{title}{Human decisions and machine predictions}.
\newblock \emph{\bibinfo{journal}{The Quarterly Journal of Economics}} \textbf{\bibinfo{volume}{133}}, \bibinfo{pages}{237--293} (\bibinfo{year}{2018}).

\bibitem{rotaru2022event}
\bibinfo{author}{Rotaru, V.}, \bibinfo{author}{Huang, Y.}, \bibinfo{author}{Li, T.}, \bibinfo{author}{Evans, J.} \& \bibinfo{author}{Chattopadhyay, I.}
\newblock \bibinfo{title}{Event-level prediction of urban crime reveals a signature of enforcement bias in us cities}.
\newblock \emph{\bibinfo{journal}{Nature human behaviour}} \textbf{\bibinfo{volume}{6}}, \bibinfo{pages}{1056--1068} (\bibinfo{year}{2022}).

\bibitem{rahwan2018society}
\bibinfo{author}{Rahwan, I.}
\newblock \bibinfo{title}{{Society-in-the-loop: programming the algorithmic social contract}}.
\newblock \emph{\bibinfo{journal}{Ethics and Information Technology}} \textbf{\bibinfo{volume}{20}}, \bibinfo{pages}{5--14} (\bibinfo{year}{2018}).

\bibitem{kobis2022promise}
\bibinfo{author}{K{\"o}bis, N.}, \bibinfo{author}{Starke, C.} \& \bibinfo{author}{Rahwan, I.}
\newblock \bibinfo{title}{The promise and perils of using artificial intelligence to fight corruption}.
\newblock \emph{\bibinfo{journal}{Nature Machine Intelligence}} \textbf{\bibinfo{volume}{4}}, \bibinfo{pages}{418--424} (\bibinfo{year}{2022}).

\bibitem{Daneshjou:2022}
\bibinfo{author}{Daneshjou, R.} \emph{et~al.}
\newblock \bibinfo{title}{Disparities in dermatology ai performance on a diverse, curated clinical image set}.
\newblock \emph{\bibinfo{journal}{Science Advances}} \textbf{\bibinfo{volume}{8}}, \bibinfo{pages}{eabq6147} (\bibinfo{year}{2022}).

\bibitem{cowgill2020biased}
\bibinfo{author}{Cowgill, B.} \emph{et~al.}
\newblock \bibinfo{title}{Biased programmers? or biased data? a field experiment in operationalizing {AI} ethics}.
\newblock In \emph{\bibinfo{booktitle}{Proceedings of the 21st ACM Conference on Economics and Computation}}, \bibinfo{pages}{679--681} (\bibinfo{year}{2020}).

\bibitem{barocas2016big}
\bibinfo{author}{Barocas, S.} \& \bibinfo{author}{Selbst, A.~D.}
\newblock \bibinfo{title}{Big data's disparate impact}.
\newblock \emph{\bibinfo{journal}{California Law Review}} \textbf{\bibinfo{volume}{104}}, \bibinfo{pages}{671} (\bibinfo{year}{2016}).

\bibitem{crawford2016there}
\bibinfo{author}{Crawford, K.} \& \bibinfo{author}{Calo, R.}
\newblock \bibinfo{title}{There is a blind spot in ai research}.
\newblock \emph{\bibinfo{journal}{Nature News}} \textbf{\bibinfo{volume}{538}}, \bibinfo{pages}{311} (\bibinfo{year}{2016}).

\bibitem{saunders2016predictions}
\bibinfo{author}{Saunders, J.}, \bibinfo{author}{Hunt, P.} \& \bibinfo{author}{Hollywood, J.~S.}
\newblock \bibinfo{title}{Predictions put into practice: a quasi-experimental evaluation of {C}hicago’s predictive policing pilot}.
\newblock \emph{\bibinfo{journal}{Journal of Experimental Criminology}} \textbf{\bibinfo{volume}{12}}, \bibinfo{pages}{347--371} (\bibinfo{year}{2016}).

\bibitem{sweeney2013discrimination}
\bibinfo{author}{Sweeney, L.}
\newblock \bibinfo{title}{Discrimination in online ad delivery}.
\newblock \emph{\bibinfo{journal}{Communications of the ACM}} \textbf{\bibinfo{volume}{56}}, \bibinfo{pages}{44--54} (\bibinfo{year}{2013}).

\bibitem{chesney2019deep}
\bibinfo{author}{Chesney, B.} \& \bibinfo{author}{Citron, D.}
\newblock \bibinfo{title}{Deep fakes: A looming challenge for privacy, democracy, and national security}.
\newblock \emph{\bibinfo{journal}{California Law Review}} \textbf{\bibinfo{volume}{107}}, \bibinfo{pages}{1753} (\bibinfo{year}{2019}).

\bibitem{lazer2018science}
\bibinfo{author}{Lazer, D.~M.} \emph{et~al.}
\newblock \bibinfo{title}{The science of fake news}.
\newblock \emph{\bibinfo{journal}{Science}} \textbf{\bibinfo{volume}{359}}, \bibinfo{pages}{1094--1096} (\bibinfo{year}{2018}).

\bibitem{allcott2017social}
\bibinfo{author}{Allcott, H.} \& \bibinfo{author}{Gentzkow, M.}
\newblock \bibinfo{title}{Social media and fake news in the 2016 election}.
\newblock \emph{\bibinfo{journal}{Journal of economic perspectives}} \textbf{\bibinfo{volume}{31}}, \bibinfo{pages}{211--36} (\bibinfo{year}{2017}).

\bibitem{schmidt2021national}
\bibinfo{author}{Schmidt, E.} \emph{et~al.}
\newblock \bibinfo{title}{National security commission on artificial intelligence (ai)}.
\newblock \bibinfo{type}{Tech. Rep.}, \bibinfo{institution}{National Security Commission on Artificial Intelligence} (\bibinfo{year}{2021}).

\bibitem{brundage2018malicious}
\bibinfo{author}{Brundage, M.} \emph{et~al.}
\newblock \bibinfo{title}{The malicious use of artificial intelligence: Forecasting, prevention, and mitigation}.
\newblock \emph{\bibinfo{journal}{arXiv preprint arXiv:1802.07228}}  (\bibinfo{year}{2018}).

\bibitem{scharre2016autonomous}
\bibinfo{author}{Scharre, P.}
\newblock \bibinfo{title}{Autonomous weapons and operational risk} (\bibinfo{year}{2016}).

\bibitem{vinuesa2020role}
\bibinfo{author}{Vinuesa, R.} \emph{et~al.}
\newblock \bibinfo{title}{The role of artificial intelligence in achieving the sustainable development goals}.
\newblock \emph{\bibinfo{journal}{Nature Communications}} \textbf{\bibinfo{volume}{11}}, \bibinfo{pages}{1--10} (\bibinfo{year}{2020}).

\bibitem{AIindex2021}
\bibinfo{author}{Zhangm, D.} \emph{et~al.}
\newblock \bibinfo{title}{The {AI} {Index} 2021 {Annual} {Report}}.
\newblock \bibinfo{howpublished}{AI Index Steering Committee, Human-Centered AI Institute, Stanford University, Stanford, CA} (\bibinfo{year}{2021}).

\bibitem{ChinaAIPlan}
\bibinfo{title}{{China} issues guideline on artificial intelligence development}.
\newblock \bibinfo{howpublished}{\url{http://english.www.gov.cn/policies/latest_releases/2017/07/20/content_281475742458322.htm}} (\bibinfo{year}{2017}).
\newblock \bibinfo{note}{Accessed: 2022-06-06}.

\bibitem{wu2020towards}
\bibinfo{author}{Wu, F.} \emph{et~al.}
\newblock \bibinfo{title}{Towards a new generation of artificial intelligence in china}.
\newblock \emph{\bibinfo{journal}{Nature Machine Intelligence}} \textbf{\bibinfo{volume}{2}}, \bibinfo{pages}{312--316} (\bibinfo{year}{2020}).

\bibitem{NationalSecurityReport}
\bibinfo{title}{{Final} report of the national security commission on artificial intelligence}.
\newblock \bibinfo{howpublished}{\url{https://www.nscai.gov/2021-final-report/}} (\bibinfo{year}{2021}).
\newblock \bibinfo{note}{Accessed: 2022-06-06}.

\bibitem{EU_Report}
\bibinfo{title}{{Coordinated} {Plan} on {Artificial} {Intelligence}}.
\newblock \bibinfo{howpublished}{\url{https://digital-strategy.ec.europa.eu/en/library/coordinated-plan-artificial-intelligence}} (\bibinfo{year}{2018}).
\newblock \bibinfo{note}{Accessed: 2022-06-06}.

\bibitem{bostrom2017superintelligence}
\bibinfo{author}{Bostrom, N.}
\newblock \emph{\bibinfo{title}{Superintelligence}} (\bibinfo{publisher}{Dunod}, \bibinfo{year}{2017}).

\bibitem{tegmark2017life}
\bibinfo{author}{Tegmark, M.}
\newblock \emph{\bibinfo{title}{Life 3.0: Being human in the age of artificial intelligence}} (\bibinfo{publisher}{Knopf}, \bibinfo{year}{2017}).

\bibitem{zhou2006emergence}
\bibinfo{author}{Zhou, P.} \& \bibinfo{author}{Leydesdorff, L.}
\newblock \bibinfo{title}{The emergence of china as a leading nation in science}.
\newblock \emph{\bibinfo{journal}{Research policy}} \textbf{\bibinfo{volume}{35}}, \bibinfo{pages}{83--104} (\bibinfo{year}{2006}).

\bibitem{xie2014china}
\bibinfo{author}{Xie, Y.}, \bibinfo{author}{Zhang, C.} \& \bibinfo{author}{Lai, Q.}
\newblock \bibinfo{title}{China’s rise as a major contributor to science and technology}.
\newblock \emph{\bibinfo{journal}{Proceedings of the National Academy of Sciences}} \textbf{\bibinfo{volume}{111}}, \bibinfo{pages}{9437--9442} (\bibinfo{year}{2014}).

\bibitem{glanzel2008triad}
\bibinfo{author}{Gl{\"a}nzel, W.}, \bibinfo{author}{Debackere, K.} \& \bibinfo{author}{Meyer, M.}
\newblock \bibinfo{title}{‘triad’ or ‘tetrad’? on global changes in a dynamic world}.
\newblock \emph{\bibinfo{journal}{Scientometrics}} \textbf{\bibinfo{volume}{74}}, \bibinfo{pages}{71--88} (\bibinfo{year}{2008}).

\bibitem{marginson2022all}
\bibinfo{author}{Marginson, S.}
\newblock \bibinfo{title}{‘all things are in flux’: China in global science}.
\newblock \emph{\bibinfo{journal}{Higher Education}} \textbf{\bibinfo{volume}{83}}, \bibinfo{pages}{881--910} (\bibinfo{year}{2022}).

\bibitem{NSF2021}
\bibinfo{author}{NSF}.
\newblock \bibinfo{title}{{Publications Output: U.S. Trends and International Comparisons}}.
\newblock \bibinfo{howpublished}{\url{https://ncses.nsf.gov/pubs/nsb20214}} (\bibinfo{year}{2021}).
\newblock \bibinfo{note}{Accessed: 2022-06-06}.

\bibitem{NSF2022:a}
\bibinfo{author}{NSF}.
\newblock \bibinfo{title}{{The State of U.S. Science and Engineering 2022}}.
\newblock \bibinfo{howpublished}{\url{https://ncses.nsf.gov/pubs/nsb20221/executive-summary}} (\bibinfo{year}{2022}).
\newblock \bibinfo{note}{Accessed: 2022-06-06}.

\bibitem{NSF2022:b}
\bibinfo{author}{NSF}.
\newblock \bibinfo{title}{{Production and Trade of Knowledge- and Technology-Intensive Industries}}.
\newblock \bibinfo{howpublished}{\url{https://ncses.nsf.gov/pubs/nsb20226/enabling-technologies}} (\bibinfo{year}{2022}).
\newblock \bibinfo{note}{Accessed: 2022-06-06}.

\bibitem{China:initiative}
\bibinfo{author}{{The United States Department of Justice}}.
\newblock \bibinfo{title}{Information about the department of justice's {China Initiative} and a compilation of china-related prosecutions since 2018}.
\newblock \bibinfo{howpublished}{\url{https://www.justice.gov/archives/nsd/information-about-department-justice-s-china-initiative-\\and-compilation-china-related}} (\bibinfo{year}{2021}).
\newblock \bibinfo{note}{Accessed: 2022-07-07}.

\bibitem{NIH:initiative:a}
\bibinfo{author}{{Francis S. Collins}}.
\newblock \bibinfo{title}{{Letter from NIH Director Francis Collins}}.
\newblock \bibinfo{howpublished}{\url{https://www.insidehighered.com/sites/default/server_files/media/NIH\%20Foreign\%20Influence\%20Letter\%20to\%20Grantees\%2008-20-18.pdf}} (\bibinfo{year}{2018}).
\newblock \bibinfo{note}{Accessed: 2022-07-07}.

\bibitem{NIH:initiative:b}
\bibinfo{author}{{Michael Lauer}}.
\newblock \bibinfo{title}{{Foreign Interference in National Institutes of Health Funding and Grant Making Processes}}.
\newblock \bibinfo{howpublished}{\url{https://grants.nih.gov/grants/files/NIH-Foreign-Interference-Findings-2016-2018.pdf}} (\bibinfo{year}{2021}).
\newblock \bibinfo{note}{Accessed: 2022-07-07}.

\bibitem{jia2022impact}
\bibinfo{author}{Jia, R.}, \bibinfo{author}{Roberts, M.~E.}, \bibinfo{author}{Wang, Y.} \& \bibinfo{author}{Yang, E.}
\newblock \bibinfo{title}{The impact of us-china tensions on us science}.
\newblock \bibinfo{type}{Tech. Rep.}, \bibinfo{institution}{National Bureau of Economic Research} (\bibinfo{year}{2022}).

\bibitem{chinesedocument:2020}
\bibinfo{author}{{Ministry of Science and Technology of the People's Republic of China}}.
\newblock \bibinfo{title}{Opinions on further strengthening the protection of intellectual property rights}.
\newblock \bibinfo{howpublished}{\url{https://www.most.gov.cn/xxgk/xinxifenlei/fdzdgknr/fgzc/gfxwj/gfxwj2020/202002/t20200223_151781.html}} (\bibinfo{year}{2020}).
\newblock \bibinfo{note}{Accessed April 10, 2023}.

\bibitem{zhong2022china}
\bibinfo{author}{Zhong, B.}, \bibinfo{author}{Liu, X.}, \bibinfo{author}{Zhan, Z.} \emph{et~al.}
\newblock \bibinfo{title}{China: reform research-evaluation criteria}.
\newblock \emph{\bibinfo{journal}{Nature}} \textbf{\bibinfo{volume}{602}}, \bibinfo{pages}{386--386} (\bibinfo{year}{2022}).

\bibitem{sinha2015overview}
\bibinfo{author}{Sinha, A.} \emph{et~al.}
\newblock \bibinfo{title}{An overview of {M}icrosoft {A}cademic {S}ervice ({MAS}) and applications}.
\newblock In \emph{\bibinfo{booktitle}{Proceedings of the 24th international conference on {World Wide Web}}}, \bibinfo{pages}{243--246} (\bibinfo{year}{2015}).

\bibitem{venkatraman2010conventions}
\bibinfo{author}{Venkatraman, V.}, \bibinfo{author}{Arzbaecher, R.}, \bibinfo{author}{Maru{\v{s}}i{\'c}, M.}, \bibinfo{author}{Maru{\v{s}}i{\'c}, A.} \& \bibinfo{author}{Marusic, A.}
\newblock \bibinfo{title}{Conventions of scientific authorship}.
\newblock \emph{\bibinfo{journal}{Science}} \textbf{\bibinfo{volume}{12}} (\bibinfo{year}{2010}).

\bibitem{alshebli2022Beijing}
\bibinfo{author}{AlShebli, B.~K.} \emph{et~al.}
\newblock \bibinfo{title}{Beijing's central role in global artificial intelligence research}.
\newblock \emph{\bibinfo{journal}{Scientific Reports}}  (\bibinfo{year}{2022}).

\bibitem{uzzi2013atypical}
\bibinfo{author}{Uzzi, B.}, \bibinfo{author}{Mukherjee, S.}, \bibinfo{author}{Stringer, M.} \& \bibinfo{author}{Jones, B.}
\newblock \bibinfo{title}{Atypical combinations and scientific impact}.
\newblock \emph{\bibinfo{journal}{Science}} \textbf{\bibinfo{volume}{342}}, \bibinfo{pages}{468--472} (\bibinfo{year}{2013}).

\bibitem{shi2023surprising}
\bibinfo{author}{Shi, F.} \& \bibinfo{author}{Evans, J.}
\newblock \bibinfo{title}{Surprising combinations of research contents and contexts are related to impact and emerge with scientific outsiders from distant disciplines}.
\newblock \emph{\bibinfo{journal}{Nature Communications}} \textbf{\bibinfo{volume}{14}}, \bibinfo{pages}{1641} (\bibinfo{year}{2023}).

\bibitem{wang2019early}
\bibinfo{author}{Wang, Y.}, \bibinfo{author}{Jones, B.~F.} \& \bibinfo{author}{Wang, D.}
\newblock \bibinfo{title}{Early-career setback and future career impact}.
\newblock \emph{\bibinfo{journal}{Nature communications}} \textbf{\bibinfo{volume}{10}}, \bibinfo{pages}{4331} (\bibinfo{year}{2019}).

\bibitem{ye2017nationality}
\bibinfo{author}{Ye, J.} \emph{et~al.}
\newblock \bibinfo{title}{Nationality classification using name embeddings}.
\newblock In \emph{\bibinfo{booktitle}{Proceedings of the 2017 ACM on Conference on Information and Knowledge Management}}, \bibinfo{pages}{1897--1906} (\bibinfo{year}{2017}).

\bibitem{alshebli2018preeminence}
\bibinfo{author}{AlShebli, B.~K.}, \bibinfo{author}{Rahwan, T.} \& \bibinfo{author}{Woon, W.~L.}
\newblock \bibinfo{title}{The preeminence of ethnic diversity in scientific collaboration}.
\newblock \emph{\bibinfo{journal}{Nature communications}} \textbf{\bibinfo{volume}{9}}, \bibinfo{pages}{1--10} (\bibinfo{year}{2018}).

\bibitem{ghosh2021fair}
\bibinfo{author}{Ghosh, A.}, \bibinfo{author}{Dutt, R.} \& \bibinfo{author}{Wilson, C.}
\newblock \bibinfo{title}{When fair ranking meets uncertain inference}.
\newblock In \emph{\bibinfo{booktitle}{Proceedings of the 44th Int. ACM SIGIR Conf. on Research \& Development in Information Retrieval}}, \bibinfo{pages}{1033--1043} (\bibinfo{year}{2021}).

\bibitem{zeina2020gender}
\bibinfo{author}{Zeina, M.}, \bibinfo{author}{Balston, A.}, \bibinfo{author}{Banerjee, A.} \& \bibinfo{author}{Woolf, K.}
\newblock \bibinfo{title}{Gender and ethnic differences in publication of bmj letters to the editor: an observational study using machine learning}.
\newblock \emph{\bibinfo{journal}{BMJ open}} \textbf{\bibinfo{volume}{10}}, \bibinfo{pages}{e037269} (\bibinfo{year}{2020}).

\bibitem{o2022ethnic}
\bibinfo{author}{O'Brochta, W.}
\newblock \bibinfo{title}{Ethnic diversity in central government cabinets}.
\newblock \emph{\bibinfo{journal}{Politics, Groups, and Identities}} \textbf{\bibinfo{volume}{10}}, \bibinfo{pages}{189--208} (\bibinfo{year}{2022}).

\bibitem{law2022public}
\bibinfo{author}{Law, K.~K.} \& \bibinfo{author}{Zuo, L.}
\newblock \bibinfo{title}{Public concern about immigration and customer complaints against minority financial advisors}.
\newblock \emph{\bibinfo{journal}{Management Science}}  (\bibinfo{year}{2022}).

\bibitem{liu2023non}
\bibinfo{author}{Liu, F.}, \bibinfo{author}{Rahwan, T.} \& \bibinfo{author}{AlShebli, B.}
\newblock \bibinfo{title}{Non-white scientists appear on fewer editorial boards, spend more time under review, and receive fewer citations}.
\newblock \emph{\bibinfo{journal}{Proceedings of the National Academy of Sciences}} \textbf{\bibinfo{volume}{120}}, \bibinfo{pages}{e2215324120} (\bibinfo{year}{2023}).

\bibitem{jeong2014drivers}
\bibinfo{author}{Jeong, S.}, \bibinfo{author}{Choi, J.~Y.} \& \bibinfo{author}{Kim, J.-Y.}
\newblock \bibinfo{title}{On the drivers of international collaboration: The impact of informal communication, motivation, and research resources}.
\newblock \emph{\bibinfo{journal}{Science and Public Policy}} \textbf{\bibinfo{volume}{41}}, \bibinfo{pages}{520--531} (\bibinfo{year}{2014}).

\bibitem{wagner2005network}
\bibinfo{author}{Wagner, C.~S.} \& \bibinfo{author}{Leydesdorff, L.}
\newblock \bibinfo{title}{Network structure, self-organization, and the growth of international collaboration in science}.
\newblock \emph{\bibinfo{journal}{Research policy}} \textbf{\bibinfo{volume}{34}}, \bibinfo{pages}{1608--1618} (\bibinfo{year}{2005}).

\bibitem{francisco2015international}
\bibinfo{author}{Francisco, J.~S.}
\newblock \bibinfo{title}{International scientific collaborations: a key to scientific success} (\bibinfo{year}{2015}).

\bibitem{iacus2012causal}
\bibinfo{author}{Iacus, S.~M.}, \bibinfo{author}{King, G.} \& \bibinfo{author}{Porro, G.}
\newblock \bibinfo{title}{Causal inference without balance checking: Coarsened exact matching}.
\newblock \emph{\bibinfo{journal}{Political analysis}} \textbf{\bibinfo{volume}{20}}, \bibinfo{pages}{1--24} (\bibinfo{year}{2012}).

\bibitem{wang2014comment}
\bibinfo{author}{Wang, J.}, \bibinfo{author}{Mei, Y.} \& \bibinfo{author}{Hicks, D.}
\newblock \bibinfo{title}{Comment on “quantifying long-term scientific impact”}.
\newblock \emph{\bibinfo{journal}{Science}} \textbf{\bibinfo{volume}{345}}, \bibinfo{pages}{149--149} (\bibinfo{year}{2014}).

\bibitem{CORE}
\bibinfo{title}{Core rankings portal}.
\newblock \bibinfo{howpublished}{\url{https://www.core.edu.au/conference-portal}} (\bibinfo{year}{2023}).
\newblock \bibinfo{note}{Accessed: 2023-09-01}.

\bibitem{frank2019evolution}
\bibinfo{author}{Frank, M.~R.}, \bibinfo{author}{Wang, D.}, \bibinfo{author}{Cebrian, M.} \& \bibinfo{author}{Rahwan, I.}
\newblock \bibinfo{title}{The evolution of citation graphs in artificial intelligence research}.
\newblock \emph{\bibinfo{journal}{Nature Machine Intelligence}} \textbf{\bibinfo{volume}{1}}, \bibinfo{pages}{79--85} (\bibinfo{year}{2019}).

\bibitem{tran2019global}
\bibinfo{author}{Tran, B.~X.} \emph{et~al.}
\newblock \bibinfo{title}{Global evolution of research in artificial intelligence in health and medicine: a bibliometric study}.
\newblock \emph{\bibinfo{journal}{Journal of clinical medicine}} \textbf{\bibinfo{volume}{8}}, \bibinfo{pages}{360} (\bibinfo{year}{2019}).

\bibitem{martinez2021research}
\bibinfo{author}{Martinez-Plumed, F.}, \bibinfo{author}{Barredo, P.}, \bibinfo{author}{Heigeartaigh, S.~O.} \& \bibinfo{author}{Hernandez-Orallo, J.}
\newblock \bibinfo{title}{Research community dynamics behind popular ai benchmarks}.
\newblock \emph{\bibinfo{journal}{Nature Machine Intelligence}} \textbf{\bibinfo{volume}{3}}, \bibinfo{pages}{581--589} (\bibinfo{year}{2021}).

\bibitem{tang2020pace}
\bibinfo{author}{Tang, X.}, \bibinfo{author}{Li, X.}, \bibinfo{author}{Ding, Y.}, \bibinfo{author}{Song, M.} \& \bibinfo{author}{Bu, Y.}
\newblock \bibinfo{title}{The pace of artificial intelligence innovations: Speed, talent, and trial-and-error}.
\newblock \emph{\bibinfo{journal}{Journal of Informetrics}} \textbf{\bibinfo{volume}{14}}, \bibinfo{pages}{101094} (\bibinfo{year}{2020}).

\bibitem{klinger2021deep}
\bibinfo{author}{Klinger, J.}, \bibinfo{author}{Mateos-Garcia, J.} \& \bibinfo{author}{Stathoulopoulos, K.}
\newblock \bibinfo{title}{Deep learning, deep change? mapping the evolution and geography of a general purpose technology}.
\newblock \emph{\bibinfo{journal}{Scientometrics}} \textbf{\bibinfo{volume}{126}}, \bibinfo{pages}{5589--5621} (\bibinfo{year}{2021}).

\bibitem{stathoulopoulos2019gender}
\bibinfo{author}{Stathoulopoulos, K.} \& \bibinfo{author}{Mateos-Garcia, J.~C.}
\newblock \bibinfo{title}{Gender diversity in ai research}.
\newblock \emph{\bibinfo{journal}{Available at SSRN 3428240}}  (\bibinfo{year}{2019}).

\bibitem{deville2014career}
\bibinfo{author}{Deville, P.} \emph{et~al.}
\newblock \bibinfo{title}{Career on the move: Geography, stratification and scientific impact}.
\newblock \emph{\bibinfo{journal}{Scientific reports}} \textbf{\bibinfo{volume}{4}}, \bibinfo{pages}{1--7} (\bibinfo{year}{2014}).

\bibitem{james2018prediction}
\bibinfo{author}{James, C.}, \bibinfo{author}{Pappalardo, L.}, \bibinfo{author}{S{\^\i}rbu, A.} \& \bibinfo{author}{Simini, F.}
\newblock \bibinfo{title}{Prediction of next career moves from scientific profiles}.
\newblock \emph{\bibinfo{journal}{arXiv preprint arXiv:1802.04830}}  (\bibinfo{year}{2018}).

\bibitem{vaccario2021reproducing}
\bibinfo{author}{Vaccario, G.}, \bibinfo{author}{Verginer, L.} \& \bibinfo{author}{Schweitzer, F.}
\newblock \bibinfo{title}{Reproducing scientists’ mobility: a data-driven model}.
\newblock \emph{\bibinfo{journal}{Scientific reports}} \textbf{\bibinfo{volume}{11}}, \bibinfo{pages}{1--11} (\bibinfo{year}{2021}).

\bibitem{van2012science}
\bibinfo{author}{Van~Noorden, R.}
\newblock \bibinfo{title}{Science on the move}.
\newblock \emph{\bibinfo{journal}{Nature}} \textbf{\bibinfo{volume}{490}}, \bibinfo{pages}{326} (\bibinfo{year}{2012}).

\bibitem{franzoni2015international}
\bibinfo{author}{Franzoni, C.}, \bibinfo{author}{Scellato, G.} \& \bibinfo{author}{Stephan, P.}
\newblock \bibinfo{title}{International mobility of research scientists: lessons from globsci}.
\newblock In \emph{\bibinfo{booktitle}{Global mobility of research scientists}}, \bibinfo{pages}{35--65} (\bibinfo{publisher}{Elsevier}, \bibinfo{year}{2015}).

\bibitem{appelt2015factors}
\bibinfo{author}{Appelt, S.}, \bibinfo{author}{van Beuzekom, B.}, \bibinfo{author}{Galindo-Rueda, F.} \& \bibinfo{author}{de~Pinho, R.}
\newblock \bibinfo{title}{Which factors influence the international mobility of research scientists?}
\newblock In \emph{\bibinfo{booktitle}{Global mobility of research scientists}}, \bibinfo{pages}{177--213} (\bibinfo{publisher}{Elsevier}, \bibinfo{year}{2015}).

\bibitem{azoulay2017mobility}
\bibinfo{author}{Azoulay, P.}, \bibinfo{author}{Ganguli, I.} \& \bibinfo{author}{Zivin, J.~G.}
\newblock \bibinfo{title}{The mobility of elite life scientists: Professional and personal determinants}.
\newblock \emph{\bibinfo{journal}{Research Policy}} \textbf{\bibinfo{volume}{46}}, \bibinfo{pages}{573--590} (\bibinfo{year}{2017}).

\bibitem{jonkers2013research}
\bibinfo{author}{Jonkers, K.} \& \bibinfo{author}{Cruz-Castro, L.}
\newblock \bibinfo{title}{Research upon return: The effect of international mobility on scientific ties, production and impact}.
\newblock \emph{\bibinfo{journal}{Research Policy}} \textbf{\bibinfo{volume}{42}}, \bibinfo{pages}{1366--1377} (\bibinfo{year}{2013}).

\bibitem{scellato2015migrant}
\bibinfo{author}{Scellato, G.}, \bibinfo{author}{Franzoni, C.} \& \bibinfo{author}{Stephan, P.}
\newblock \bibinfo{title}{Migrant scientists and international networks}.
\newblock \emph{\bibinfo{journal}{Research Policy}} \textbf{\bibinfo{volume}{44}}, \bibinfo{pages}{108--120} (\bibinfo{year}{2015}).

\bibitem{baker2015non}
\bibinfo{author}{B{\"a}ker, A.}
\newblock \bibinfo{title}{Non-tenured post-doctoral researchers’ job mobility and research output: An analysis of the role of research discipline, department size, and coauthors}.
\newblock \emph{\bibinfo{journal}{Research Policy}} \textbf{\bibinfo{volume}{44}}, \bibinfo{pages}{634--650} (\bibinfo{year}{2015}).

\bibitem{petersen2018multiscale}
\bibinfo{author}{Petersen, A.~M.}
\newblock \bibinfo{title}{Multiscale impact of researcher mobility}.
\newblock \emph{\bibinfo{journal}{Journal of the Royal Society Interface}} \textbf{\bibinfo{volume}{15}}, \bibinfo{pages}{20180580} (\bibinfo{year}{2018}).

\bibitem{doria2017quantifying}
\bibinfo{author}{Doria~Arrieta, O.~A.}, \bibinfo{author}{Pammolli, F.} \& \bibinfo{author}{Petersen, A.~M.}
\newblock \bibinfo{title}{Quantifying the negative impact of brain drain on the integration of {E}uropean science}.
\newblock \emph{\bibinfo{journal}{Science advances}} \textbf{\bibinfo{volume}{3}}, \bibinfo{pages}{e1602232} (\bibinfo{year}{2017}).

\bibitem{wagner2018openness}
\bibinfo{author}{Wagner, C.~S.}, \bibinfo{author}{Whetsell, T.}, \bibinfo{author}{Baas, J.} \& \bibinfo{author}{Jonkers, K.}
\newblock \bibinfo{title}{Openness and impact of leading scientific countries}.
\newblock \emph{\bibinfo{journal}{Frontiers in research metrics and analytics}} \textbf{\bibinfo{volume}{3}}, \bibinfo{pages}{10} (\bibinfo{year}{2018}).

\bibitem{chinchilla2018global}
\bibinfo{author}{Chinchilla-Rodr{\'\i}guez, Z.} \emph{et~al.}
\newblock \bibinfo{title}{A global comparison of scientific mobility and collaboration according to national scientific capacities}.
\newblock \emph{\bibinfo{journal}{Frontiers in research metrics and analytics}} \textbf{\bibinfo{volume}{3}}, \bibinfo{pages}{17} (\bibinfo{year}{2018}).

\bibitem{verginer2020cities}
\bibinfo{author}{Verginer, L.} \& \bibinfo{author}{Riccaboni, M.}
\newblock \bibinfo{title}{Cities and countries in the global scientist mobility network}.
\newblock \emph{\bibinfo{journal}{Applied Network Science}} \textbf{\bibinfo{volume}{5}}, \bibinfo{pages}{1--16} (\bibinfo{year}{2020}).

\bibitem{dyachenko2017internal}
\bibinfo{author}{Dyachenko, E.~L.}
\newblock \bibinfo{title}{Internal migration of scientists in russia and the usa: the case of physicists}.
\newblock \emph{\bibinfo{journal}{Scientometrics}} \textbf{\bibinfo{volume}{113}}, \bibinfo{pages}{105--122} (\bibinfo{year}{2017}).

\bibitem{huang2014mobility}
\bibinfo{author}{Huang, K. G.-L.} \& \bibinfo{author}{Ertug, G.}
\newblock \bibinfo{title}{Mobility, retention and productivity of genomics scientists in the united states}.
\newblock \emph{\bibinfo{journal}{Nature Biotechnology}} \textbf{\bibinfo{volume}{32}}, \bibinfo{pages}{953--958} (\bibinfo{year}{2014}).

\bibitem{slavova2016learning}
\bibinfo{author}{Slavova, K.}, \bibinfo{author}{Fosfuri, A.} \& \bibinfo{author}{De~Castro, J.~O.}
\newblock \bibinfo{title}{Learning by hiring: The effects of scientists’ inbound mobility on research performance in academia}.
\newblock \emph{\bibinfo{journal}{Organization Science}} \textbf{\bibinfo{volume}{27}}, \bibinfo{pages}{72--89} (\bibinfo{year}{2016}).

\bibitem{levine2006research}
\bibinfo{author}{Levine, A.~D.}
\newblock \bibinfo{title}{Research policy and the mobility of us stem cell scientists}.
\newblock \emph{\bibinfo{journal}{Nature biotechnology}} \textbf{\bibinfo{volume}{24}}, \bibinfo{pages}{865--866} (\bibinfo{year}{2006}).

\bibitem{glanzel2001national}
\bibinfo{author}{Gl{\"a}nzel, W.}
\newblock \bibinfo{title}{National characteristics in international scientific co-authorship relations}.
\newblock \emph{\bibinfo{journal}{Scientometrics}} \textbf{\bibinfo{volume}{51}}, \bibinfo{pages}{69--115} (\bibinfo{year}{2001}).

\bibitem{katz1997much}
\bibinfo{author}{Katz, J.} \& \bibinfo{author}{Hicks, D.}
\newblock \bibinfo{title}{How much is a collaboration worth? a calibrated bibliometric model}.
\newblock \emph{\bibinfo{journal}{Scientometrics}} \textbf{\bibinfo{volume}{40}}, \bibinfo{pages}{541--554} (\bibinfo{year}{1997}).

\bibitem{aksnes2003characteristics}
\bibinfo{author}{Aksnes, D.~W.}
\newblock \bibinfo{title}{Characteristics of highly cited papers}.
\newblock \emph{\bibinfo{journal}{Research evaluation}} \textbf{\bibinfo{volume}{12}}, \bibinfo{pages}{159--170} (\bibinfo{year}{2003}).

\bibitem{persson2010highly}
\bibinfo{author}{Persson, O.}
\newblock \bibinfo{title}{Are highly cited papers more international?}
\newblock \emph{\bibinfo{journal}{Scientometrics}} \textbf{\bibinfo{volume}{83}}, \bibinfo{pages}{397--401} (\bibinfo{year}{2010}).

\bibitem{kwiek2021large}
\bibinfo{author}{Kwiek, M.}
\newblock \bibinfo{title}{What large-scale publication and citation data tell us about international research collaboration in {E}urope: Changing national patterns in global contexts}.
\newblock \emph{\bibinfo{journal}{Studies in Higher Education}} \textbf{\bibinfo{volume}{46}}, \bibinfo{pages}{2629--2649} (\bibinfo{year}{2021}).

\bibitem{van2004sleeping}
\bibinfo{author}{Van~Raan, A.~F.}
\newblock \bibinfo{title}{Sleeping beauties in science}.
\newblock \emph{\bibinfo{journal}{Scientometrics}} \textbf{\bibinfo{volume}{59}}, \bibinfo{pages}{467--472} (\bibinfo{year}{2004}).

\bibitem{sinatra2016quantifying}
\bibinfo{author}{Sinatra, R.}, \bibinfo{author}{Wang, D.}, \bibinfo{author}{Deville, P.}, \bibinfo{author}{Song, C.} \& \bibinfo{author}{Barab{\'a}si, A.-L.}
\newblock \bibinfo{title}{Quantifying the evolution of individual scientific impact}.
\newblock \emph{\bibinfo{journal}{Science}} \textbf{\bibinfo{volume}{354}}, \bibinfo{pages}{aaf5239} (\bibinfo{year}{2016}).

\bibitem{CSCD}
\bibinfo{author}{{Clarivate Analytics}}.
\newblock \bibinfo{title}{{Chinese Science Citation Database (CSCD)}} (\bibinfo{year}{2024}).
\newblock \urlprefix\url{https://wokinfo.com/products_tools/multidisciplinary/cscd/}.
\newblock \bibinfo{note}{Accessed: 2024-11-06}.

\bibitem{CSSCI}
\bibinfo{author}{{Nanjing University}}.
\newblock \bibinfo{title}{{Chinese Social Sciences Citation Index (CSSCI)}} (\bibinfo{year}{2024}).
\newblock \urlprefix\url{http://cssci.nju.edu.cn/}.
\newblock \bibinfo{note}{Accessed: 2024-11-06}.

\bibitem{galanos2022nomadic}
\bibinfo{author}{Galanos, V.}
\newblock \bibinfo{title}{Nomadic artificial intelligence and royal research councils: Curiosity-driven research against imperatives implying imperialism}.
\newblock In \emph{\bibinfo{booktitle}{The Global Politics of Artificial Intelligence}}, \bibinfo{pages}{173--208} (\bibinfo{publisher}{Chapman and Hall/CRC}, \bibinfo{year}{2022}).

\bibitem{galanos2023expectations}
\bibinfo{author}{Galanos, V.}
\newblock \bibinfo{title}{Expectations and expertise in artificial intelligence: specialist views and historical perspectives on conceptualisation, promise, and funding}  (\bibinfo{year}{2023}).

\bibitem{ding2022dueling}
\bibinfo{author}{Ding, J.}
\newblock \bibinfo{title}{Dueling perspectives in ai and us--china relations: Technonationalism vs. technoglobalism}  (\bibinfo{year}{2022}).

\bibitem{daniels2022knowledge}
\bibinfo{author}{Daniels, M.} \& \bibinfo{author}{Krige, J.}
\newblock \emph{\bibinfo{title}{Knowledge regulation and national security in postwar America}} (\bibinfo{publisher}{University of Chicago Press}, \bibinfo{year}{2022}).

\end{thebibliography}
\bibliographystyle{naturemag}

\section*{Author Contributions}
B.A., J.A.E. and T.R. conceived the study; B.A., J.A.E. and T.R. designed the research; B.A. and S.A.M. collected and processed the data; B.A., S.A.M. and T.R. designed and produced the visualizations; B.A. and T.R. wrote the main manuscript text; S.A.M. and J.A.E. contributed to writing; all authors reviewed the manuscript.

\section*{Competing interests}
The authors declare no competing interests.

\clearpage

\clearpage

\end{document}


\baselineskip24pt

\maketitle 
\vspace{-1cm}
\clearpage
{\noindent\Large \textbf{Supplementary Figures}}

\begin{figure}[H]
\centering
\includegraphics[width=\textwidth]{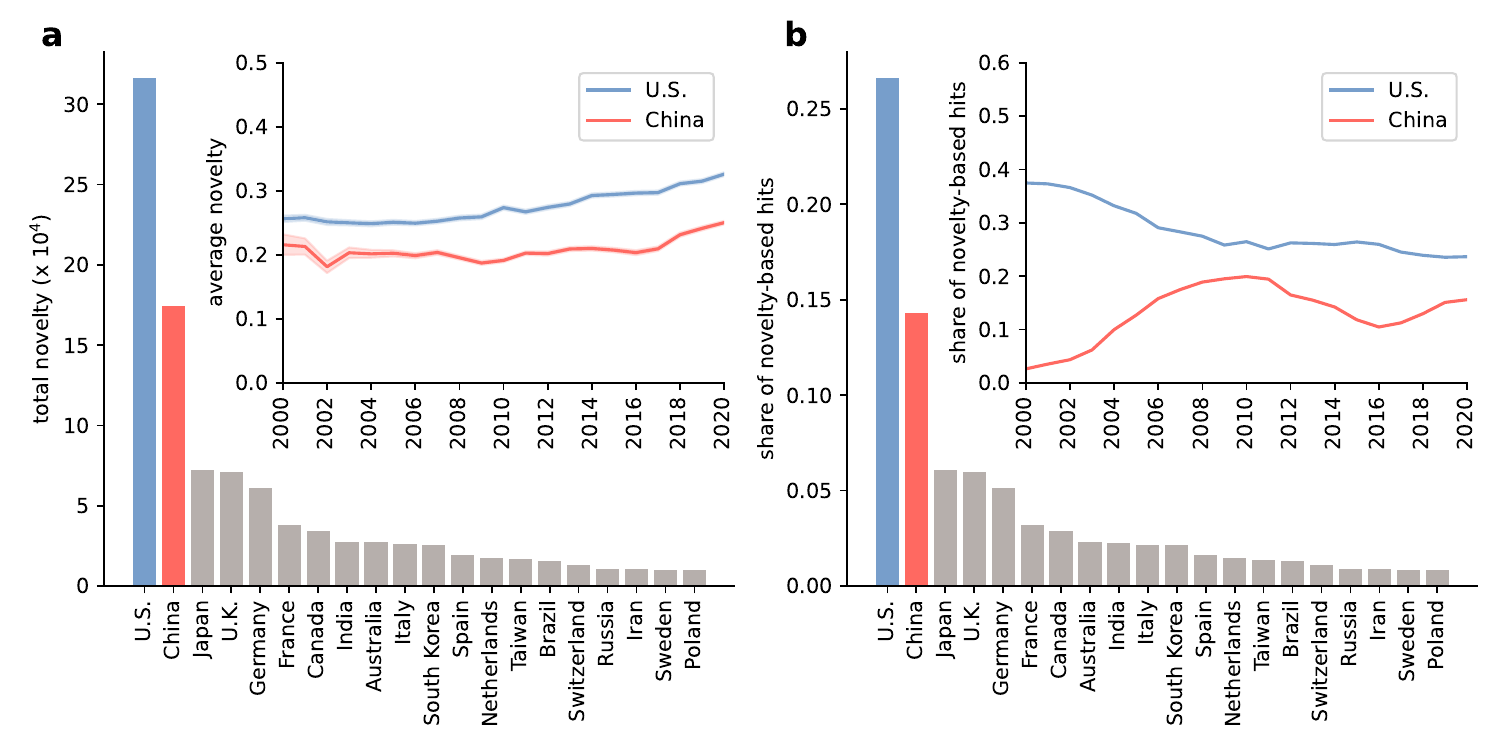}
\caption{\textbf{Content novelty analysis.} \textbf{a,} Main plot: The 20 countries that garner the highest total content novelty in AI research. Inset: The average content novelty of AI papers in the leading two countries (U.S.\ and China) over time. \textbf{b,} Main plot: The 20 countries
with the greatest share of hits based on content novelty. Inset: The share of hit AI papers based on content novelty in the
leading two countries (U.S.\ and China) over time.}
\label{fig_supplementary:novelty_evans}
\end{figure}
\clearpage

\begin{figure}[H]
\centering
\includegraphics[width=\textwidth]{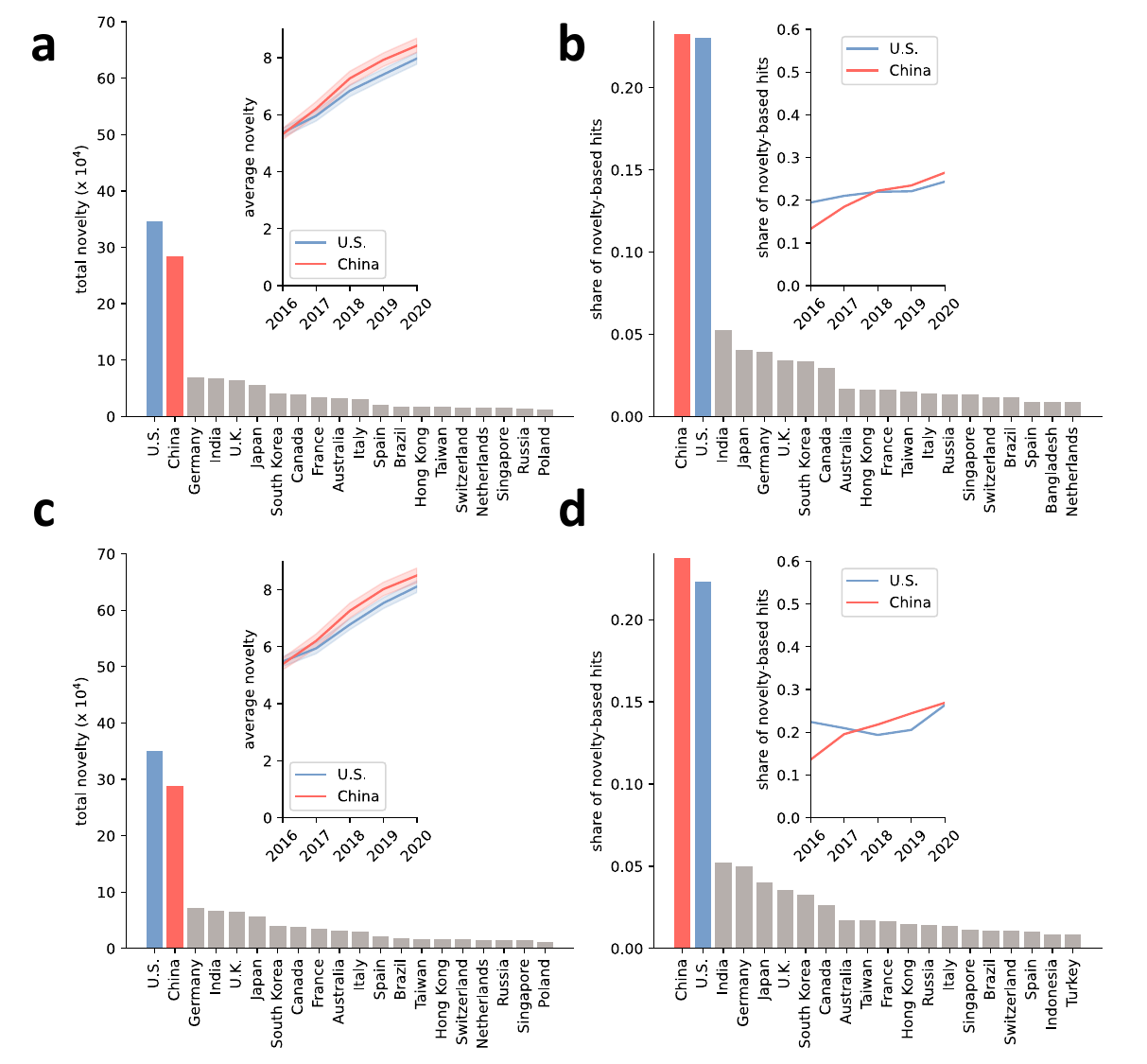}
\caption{\textbf{Context novelty analysis after excluding relatively new venues.} Analyzing how context novelty in AI papers published in [2016,2020] would change as a result of excluding references published in journals that are established after 2015. \textbf{a}, Main plot: the 20 countries with the highest context novelty in AI papers published in [2016,2020] (before excluding references). Inset: The average result over time for the two leading countries, U.S. and China. \textbf{b}, Main plot: The 20 countries with the greatest share of AI hits based on impact in [2016,2020] (before excluding references). Inset: The average result over time for the two leading countries, U.S. and China. \textbf{c} and \textbf{d}, The same as \textbf{(a)} and \textbf{(b)}, respectively, but after excluding references published in journals that are established after 2015.}
\label{fig_supplementary:novelty_robustness}
\end{figure}
\clearpage

\begin{figure}[h!]
    \centering
    \begin{minipage}{0.85\textwidth}
        \centering
        \includegraphics[width=\linewidth]{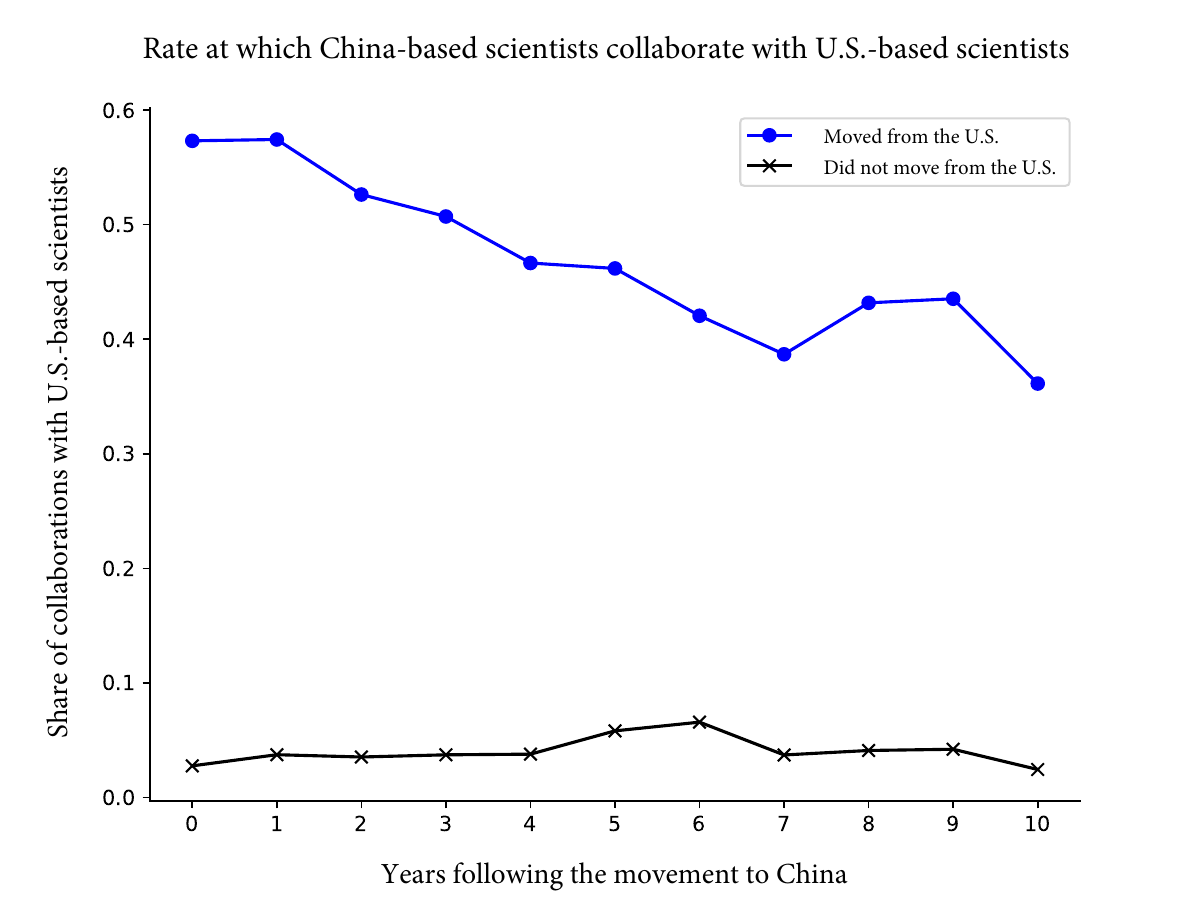}
    \end{minipage}
    \begin{minipage}{0.85\textwidth}
        \centering
        \includegraphics[width=\linewidth]{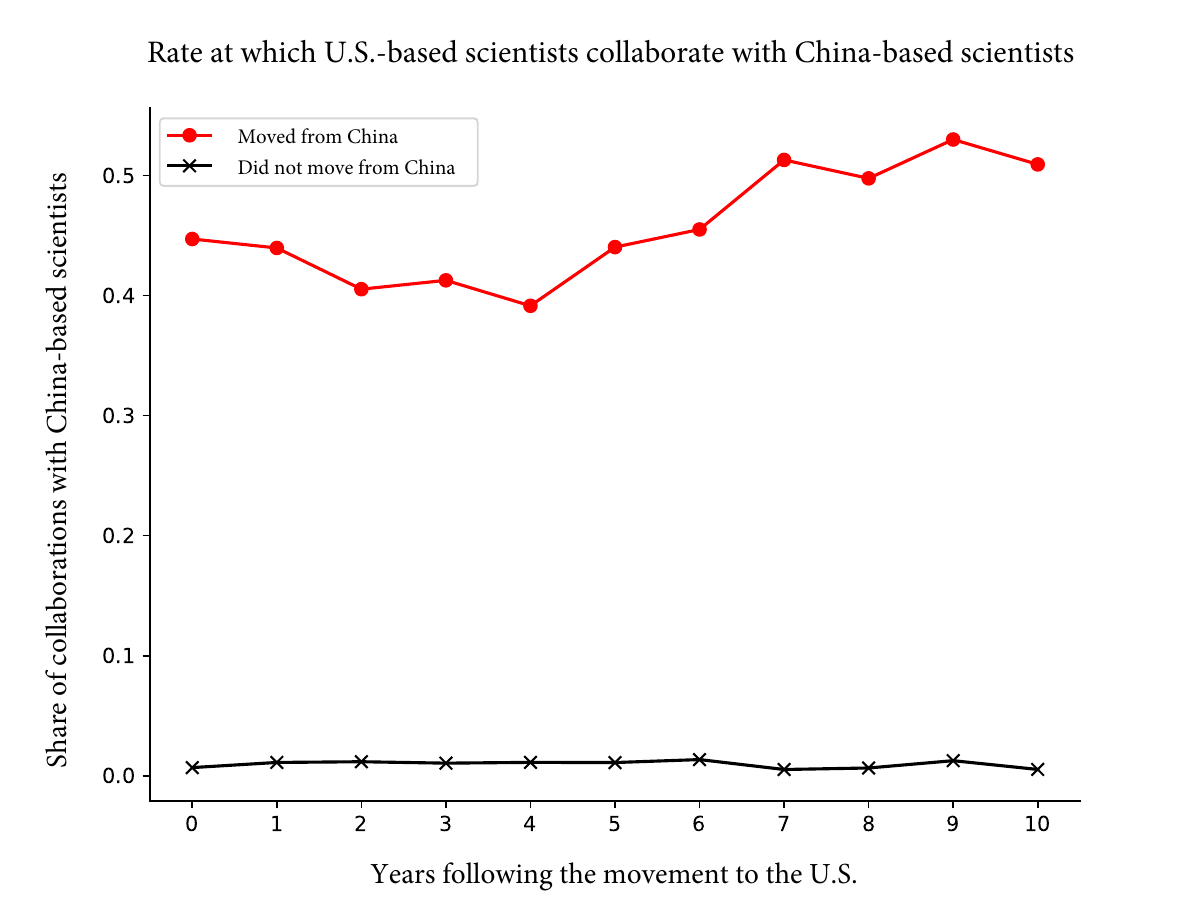}
    \end{minipage}
    \caption{{\textbf{Collaboration rate with the country of origin over time.} The same as Figure 2f, but over the 10 years that followed the move.}}
    \label{fig:sidebyside}
\end{figure}
\clearpage

\begin{figure}[H]
    \centering    {\includegraphics[height=0.6\textwidth]{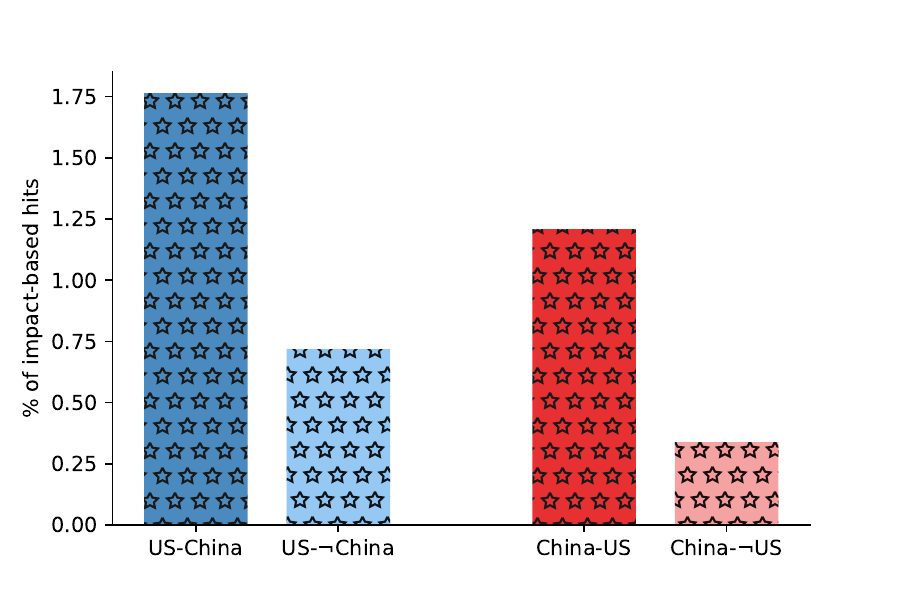}}
    \caption{\textbf{Impact-based AI hit rates.} Comparing the rate of impact-based AI hits for the four types of AI papers: (i) U.S.-based papers produced in collaboration with China, $(\mathit{US}, \mathit{China})$; (ii) U.S.-based papers produced without China, $(\mathit{US}, \neg\mathit{China})$; (iii) China-based papers produced in collaboration with the U.S., $(\mathit{China}, \mathit{US})$; (iv) China-based papers produced without the U.S., $(\mathit{China}, \neg\mathit{US})$.}    \label{fig_supplementary:collab_premium_impact_hits}
\end{figure}
\clearpage

\begin{figure}[H]
\centering
\includegraphics[width=0.94\textwidth]{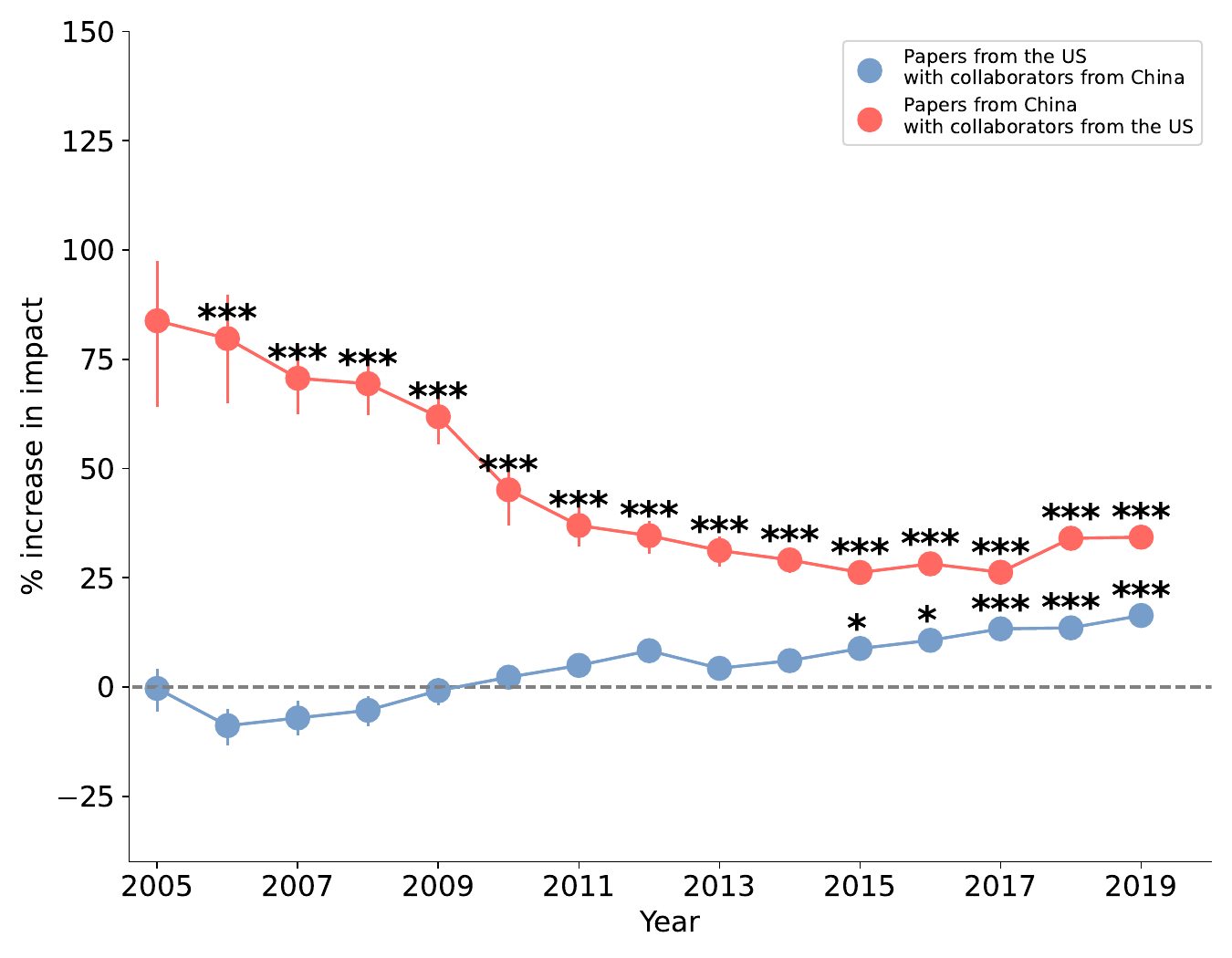}
\caption{{\textbf{Within-subject analysis of U.S.-China collaborations.} The same as Figure 3d, but controlling for the last author. That is, we compare papers that involve U.S.-China collaborations to those produced by one country without the other, but the comparison is now performed among papers that have the same last-author. When performing this comparison, we allow for up to two years difference in publication year. Moreover, we bin the sizes of teams that involve five or more authors, using the following bins: [5,6], [7,9], [10,$\infty$].}}
\label{fig_supplementary:withinperson}
\end{figure}
\clearpage

\begin{figure}[H]
\centering
\includegraphics[width=\textwidth]{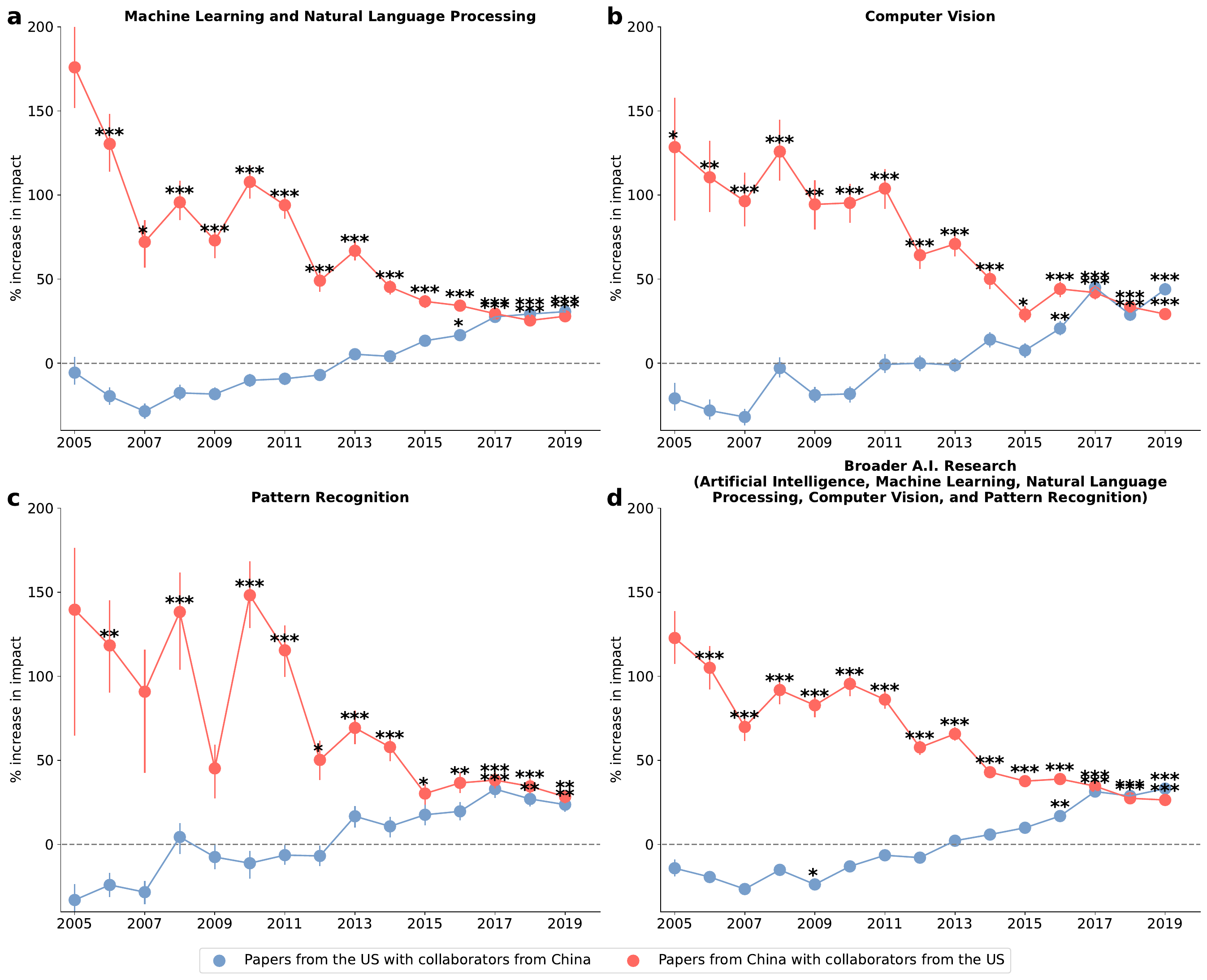}
\caption{{\textbf{Other AI-related subfields} The same as Figure 3d, but examining the following AI-related areas: (\textbf{a}) Machine Learning and Natural Language Processing, (\textbf{b}) Computer Vision, (\textbf{c}) Pattern Recognition, and (\textbf{d}) a broader definition of AI research that encompasses the three aforementioned areas (\textbf{a}, \textbf{b}, and \textbf{c}) in addition to the area of ``AI''.}}
\label{fig_supplementary:withinperson}
\end{figure}
\clearpage

\begin{figure}[H]
\centering
\includegraphics[width=0.94\textwidth]{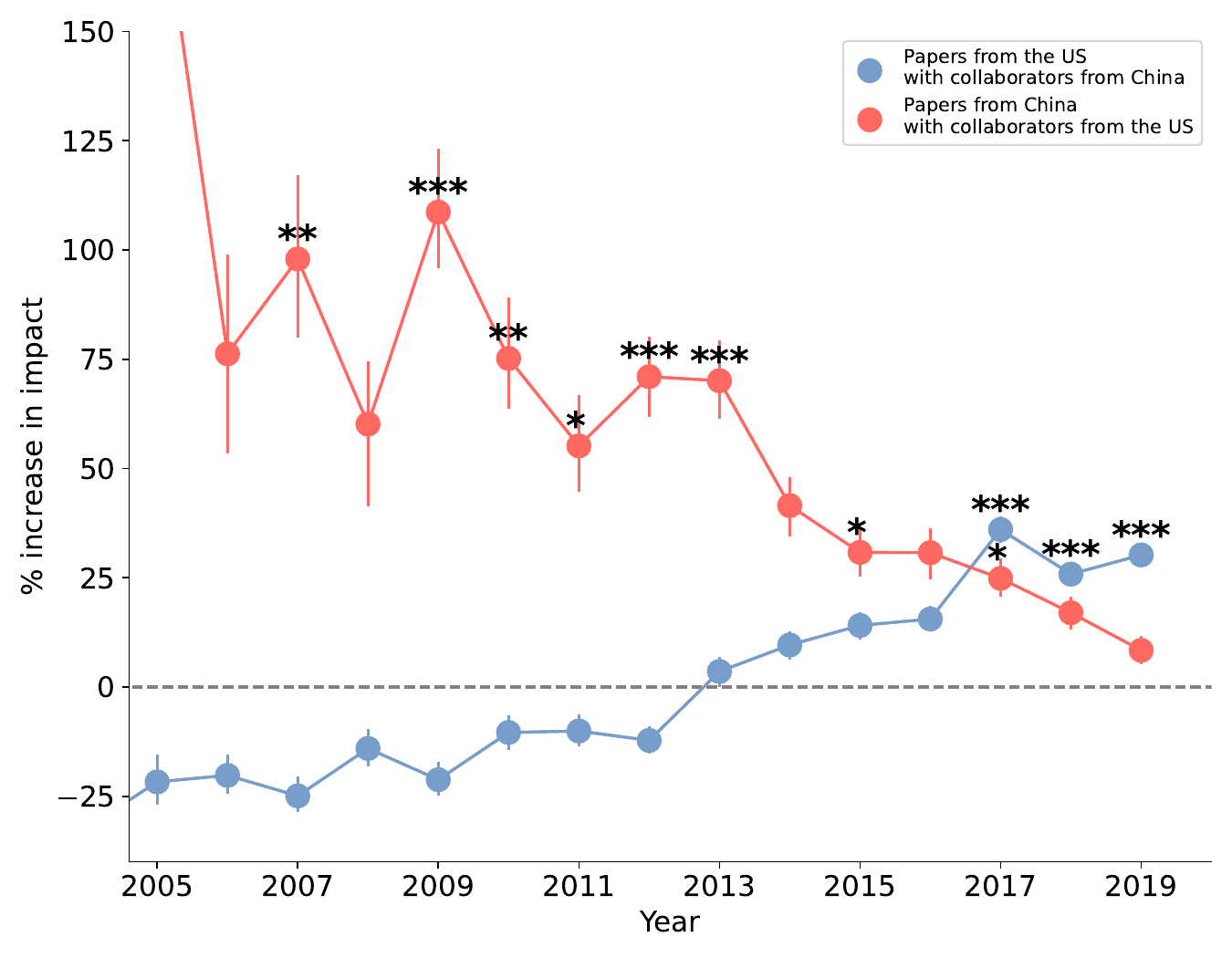}
\caption{\textbf{Alternative Collaboration Criteria.} The same as Figure~3d, but only focusing on papers where the last author is in one country and the first author is in the other.}
\label{fig_supplementary:firstAuthor}
\end{figure}
\clearpage

\begin{figure}[H]
\centering
\includegraphics[width=0.94\textwidth]{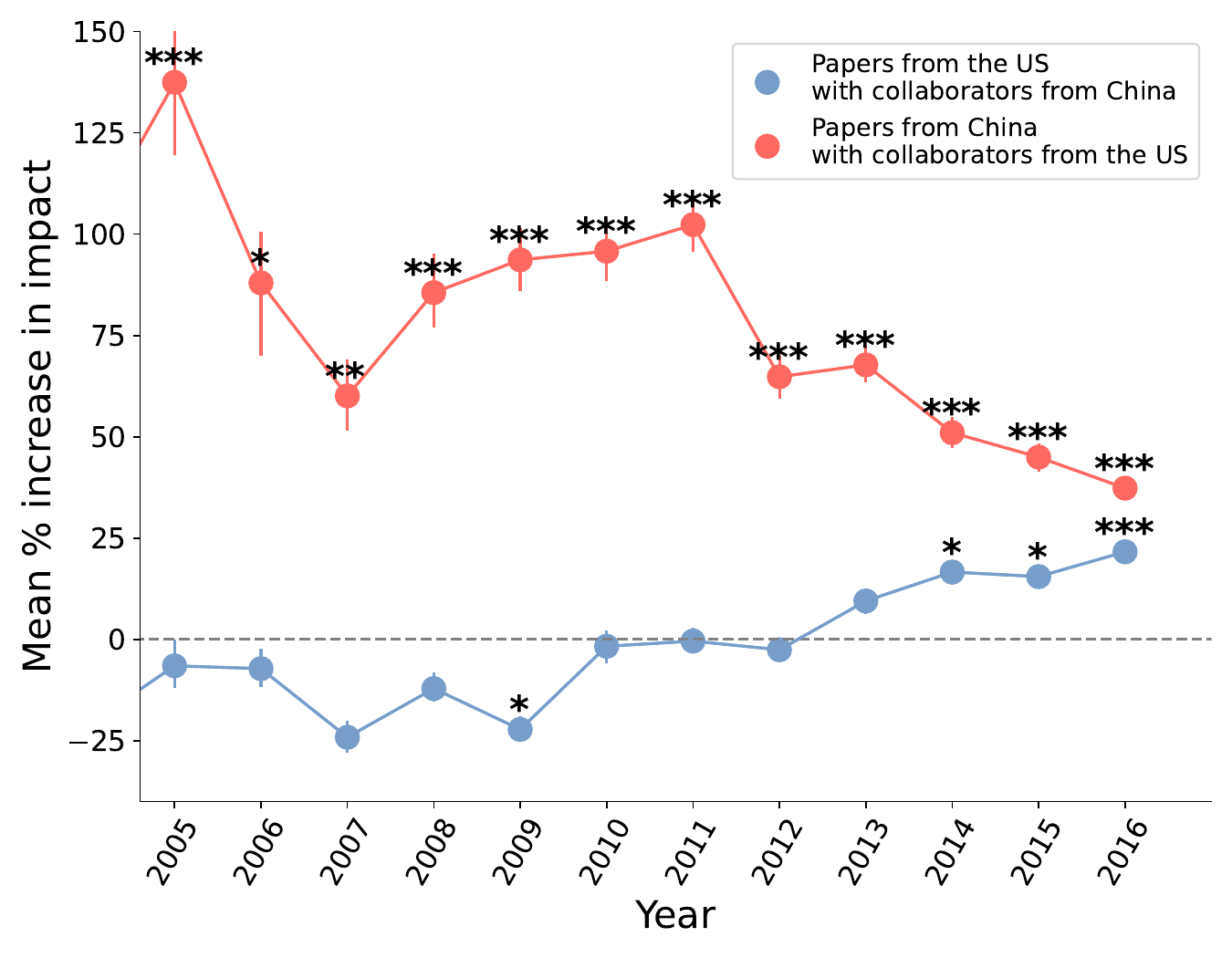}
\caption{\textbf{Citations within five years.}The same as Figure~3d, but for the number of citations received within five years, instead of two years, post publication.}
\label{fig_supplementary:impact_c5}
\end{figure}
\clearpage

\begin{figure}[H]
    \centering
    {\includegraphics[width=\textwidth]{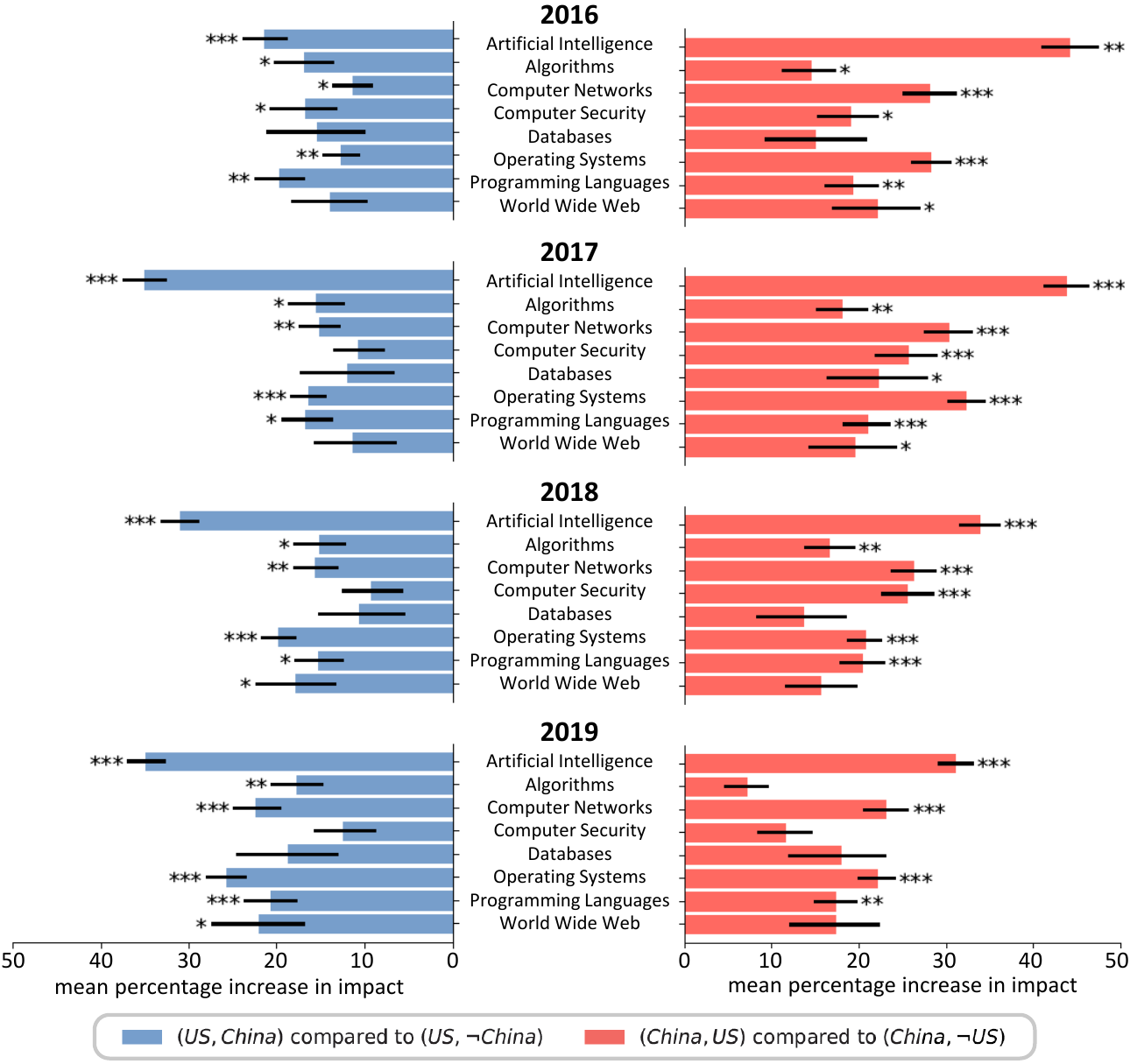}}
    \caption{\textbf{Examining U.S.-China collaborations across Computer Science subfields.} Analyzing the four types of papers used in Figure 3a, i.e., $(\mathit{US}, \mathit{China})$, $(\mathit{US}, \neg\mathit{China})$, $(\mathit{China}, \mathit{US})$, and $(\mathit{China}, \neg\mathit{US})$ using Coarsened Exact Matching (CEM) to quantify the percentage increase in impact of $(\mathit{US}, \mathit{China})$ compared to $(\mathit{US}, \neg\mathit{China})$, as well as  $(\mathit{China}, \mathit{US})$ compared to $(\mathit{China}, \neg\mathit{US})$. However, unlike Figure 3a, the comparison is now done in different fields of Computer Science (rather than AI only), for the years 2016 to 2019 (inclusive).}
    \label{fig:otherFields}

\end{figure}
\clearpage

\begin{figure}[H]
    \centering
    {\includegraphics[height=0.87\textheight]{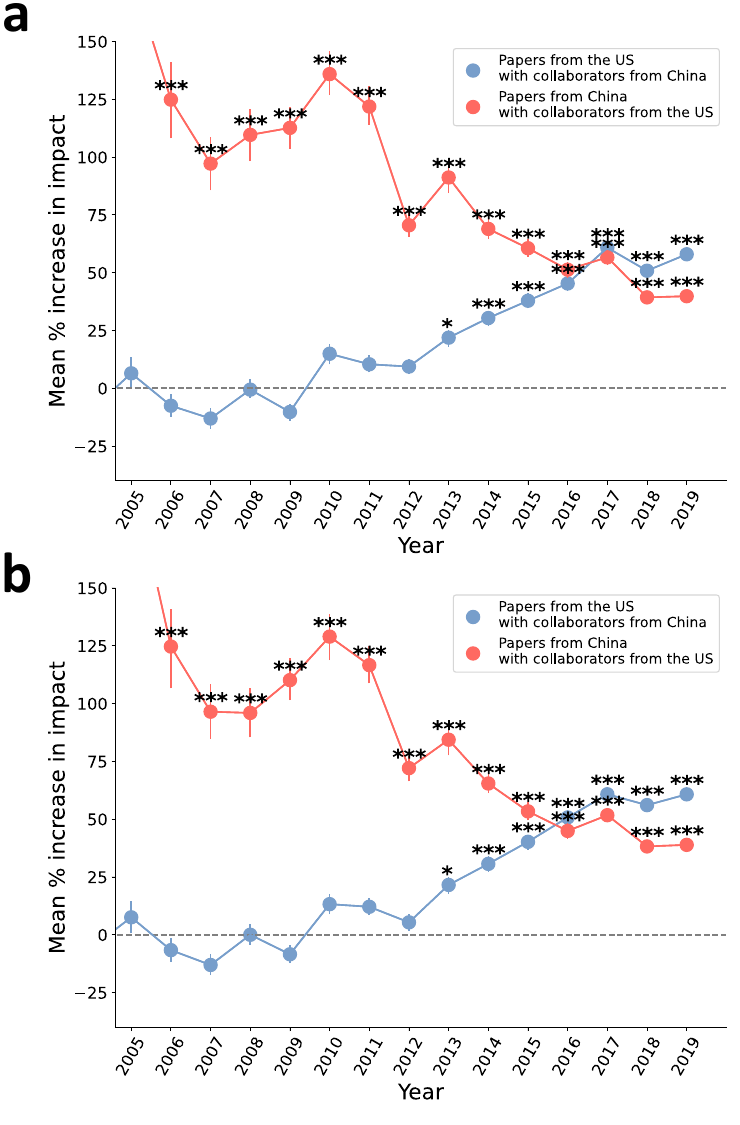}}
    \caption{\textbf{Excluding home citation bias.}The same as Figure~3d, but after removing ``home citations" based on last author affiliation (\textbf{a}), and based on any-author affiliation (\textbf{b}).}
    \label{fig:homecitations}
\end{figure}
\clearpage

\begin{figure}[H]
    \centering
    {\includegraphics[width=\textwidth]{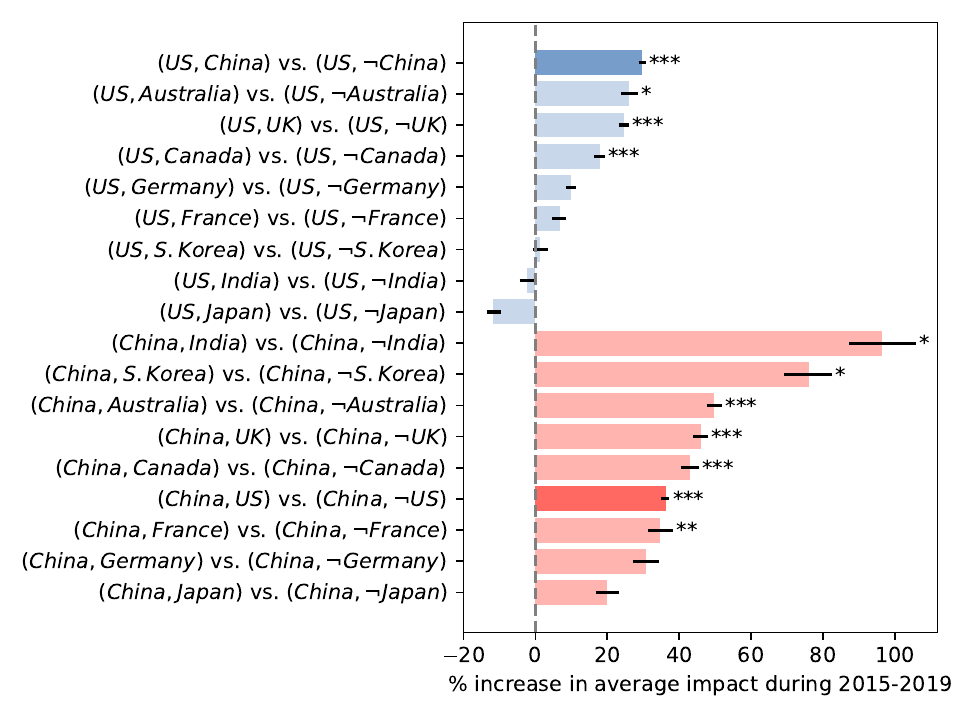}}
    \caption{\textbf{Exploring the impact when the U.S. and China collaborate with other countries.} The other countries considered in this analysis are those appearing among the 10 most productive countries as per Figure~1a. For each country, $X \neq \mathit{US}$, the figure depicts (as a blue bar) the difference in impact between the papers in $(\mathit{US}, \mathit{X})$ and those in $(\mathit{US}, \neg\mathit{X})$ over the last five years in our dataset (2015-2019). Similarly, for each country $Y \neq \mathit{China}$, the figure depicts (as a red bar) the difference in impact between the papers in $(\mathit{China}, \mathit{Y})$ and those in $(\mathit{China}, \neg\mathit{Y})$. The comparison is performed using CEM as per Figure~3d. P values are calculated using t-tests; * $p<.05$; ** $p<.01$; *** $p<.001$. }
    \label{fig:homecitations}
\end{figure}
\clearpage

{\noindent\Large \textbf{Supplementary Tables}}

\begin{table}[H]
{\fontsize{9}{9}\selectfont{
\caption{\textbf{Coarsened Exact Matching results for the migration analysis of Figure~2f.} $T'$ and $C'$ are the populations of matched treatment and matched control papers, respectively; $\mathcal{L}_1$ is the multivariate imbalance statistic~\cite{iacus2012causal}; $\mu_{T'}$ is the percentage of papers that include a China-based collaborator (first row) or a U.S.-based collaborator (second row) in $T'$; $\mu_{C'}$ is the percentage of papers that include a China-based collaborator (first row) or a U.S.-based collaborator (second row) in $C'$; a bootstrap of 95\% confidence interval ($CI_{95\%}$) is provided for $\mu_{T'}$ and $\mu_{C'}$; a t-test shows which $\delta$ values are statistically significant; see the resulting $p$-values.} 
\label{supplementarytab:migration_matching}
\begin{center}
\begin{tabular}{p{4cm}ccccccccc}
\toprule
  & $n$ & $|T'|$ & $|C'|$ & $\mathcal{L}_1$ & $\mu_{T'}$ & $\mu_{C'}$ & $CI_{95\%,T'}$ &  $CI_{95\%,C'}$  & $p$ \\
\midrule
 T: U.S.-based scientists who migrated from China\\C: U.S.-based scientists who did not migrate from China&   \hspace*{0.1cm}$4,740$ &   \hspace*{0.1cm}$1,958$ &  \hspace*{0.1cm}$1,856$ & \hspace*{0.1cm}$0.24$ &  \hspace*{0.1cm}$45.83$ & \hspace*{0.1cm}$1.24$ &  [44.22, 47.44] & [0.92, 1.55] & $<.001$ \\
 \midrule
T: China-based scientists who migrated from the U.S.\\C: China-based scientists who did not migrate from the U.S. &   \hspace*{0.1cm}$2,466$ & \hspace*{0.1cm}$700$ &  \hspace*{0.1cm}$765$ & \hspace*{0.1cm}$0.19$ &  \hspace*{0.1cm}$55.09$ & \hspace*{0.1cm}$3.01$ & [52.67, 57.52] &  [2.21, 3.82] &   $<.001$ \\
\bottomrule
\end{tabular}
\end{center}
}}
\end{table}

\clearpage

\begin{table}[H]
{\fontsize{10}{10}\selectfont{
\caption{\textbf{Coarsened Exact Matching results for U.S.-based papers produced in collaboration with China.} $T$ and $C$ are the treatment and control populations, respectively; $T'$ and $C'$ are the populations of matched treatment and matched control papers, respectively; $\mathcal{L}_1$ is  the multivariate imbalance statistic~\cite{iacus2012causal}; $\mu_{T'}$ is the mean impact (i.e., mean $c_2$) of $T'$; $\mu_{C'}$ is the mean impact of $C'$; $\delta$ is the relative impact gain of $T'$ over $C'$, i.e., $\delta = 100\times( \mu_{T'} - \mu_{C'})/\mu_{C'}$; a bootstrap of 95\% confidence interval ($CI_{95\%}$) is provided; a t-test shows which $\delta$ values are statistically significant; see the resulting $p$-values. }
\label{supplementarytab:us_china_matching}
\begin{center}
\begin{tabular}{lcccccccccc}
\toprule
 year &                  $|T|$ &                   $|C|$ &                 $|T'|$ &                  $|C'|$ &        $\mathcal{L}_1$ &                 $\mu_{T'}$ &                 $\mu_{C'}$ &                 $\delta$ &        $CI_{95\%}$ &                    $p$ \\
\midrule
  2005 &   \hspace*{0.1cm}$345$ & \hspace*{0.1cm}$35,925$ &   \hspace*{0.1cm}$314$ &  \hspace*{0.1cm}$7,165$ & \hspace*{0.1cm}$0.44$ & \hspace*{0.1cm}$5.06$ & \hspace*{0.1cm}$5.79$ & \hspace*{0.1cm}$-12.73$ &  [-17.74, -6.81] & \hspace*{0.1cm}$0.441$ \\
 2006 &   \hspace*{0.1cm}$556$ & \hspace*{0.1cm}$39,016$ &   \hspace*{0.1cm}$512$ & \hspace*{0.1cm}$10,753$ & \hspace*{0.1cm}$0.48$ & \hspace*{0.1cm}$4.15$ & \hspace*{0.1cm}$5.02$ & \hspace*{0.1cm}$-17.42$ & [-21.46, -13.06] & \hspace*{0.1cm}$0.201$ \\
 2007 &   \hspace*{0.1cm}$605$ & \hspace*{0.1cm}$41,961$ &   \hspace*{0.1cm}$531$ & \hspace*{0.1cm}$11,600$ & \hspace*{0.1cm}$0.45$ & \hspace*{0.1cm}$3.97$ & \hspace*{0.1cm}$5.35$ & \hspace*{0.1cm}$-25.86$ & [-29.50, -22.18] & \hspace*{0.1cm}$0.074$ \\
 2008 &   \hspace*{0.1cm}$820$ & \hspace*{0.1cm}$43,795$ &   \hspace*{0.1cm}$747$ & \hspace*{0.1cm}$15,791$ & \hspace*{0.1cm}$0.44$ & \hspace*{0.1cm}$4.31$ & \hspace*{0.1cm}$4.97$ & \hspace*{0.1cm}$-13.21$ &  [-16.70, -9.41] & \hspace*{0.1cm}$0.248$ \\
 2009 & \hspace*{0.1cm}$1,070$ & \hspace*{0.1cm}$44,801$ &   \hspace*{0.1cm}$983$ & \hspace*{0.1cm}$17,373$ & \hspace*{0.1cm}$0.45$ & \hspace*{0.1cm}$4.20$ & \hspace*{0.1cm}$5.34$ & \hspace*{0.1cm}$-21.38$ & [-24.41, -17.98] & \hspace*{0.1cm}$0.042$ \\
 2010 & \hspace*{0.1cm}$1,280$ & \hspace*{0.1cm}$48,163$ & \hspace*{0.1cm}$1,169$ & \hspace*{0.1cm}$22,790$ & \hspace*{0.1cm}$0.45$ & \hspace*{0.1cm}$4.69$ & \hspace*{0.1cm}$4.93$ &  \hspace*{0.1cm}$-4.93$ &   [-8.98, -1.51] & \hspace*{0.1cm}$0.697$ \\
 2011 & \hspace*{0.1cm}$1,420$ & \hspace*{0.1cm}$49,730$ & \hspace*{0.1cm}$1,301$ & \hspace*{0.1cm}$23,540$ & \hspace*{0.1cm}$0.42$ & \hspace*{0.1cm}$5.02$ & \hspace*{0.1cm}$5.21$ &  \hspace*{0.1cm}$-3.78$ &   [-6.67, -0.58] & \hspace*{0.1cm}$0.673$ \\
 2012 & \hspace*{0.1cm}$1,661$ & \hspace*{0.1cm}$53,169$ & \hspace*{0.1cm}$1,542$ & \hspace*{0.1cm}$24,835$ & \hspace*{0.1cm}$0.43$ & \hspace*{0.1cm}$5.22$ & \hspace*{0.1cm}$5.52$ &  \hspace*{0.1cm}$-5.39$ &   [-8.17, -2.76] & \hspace*{0.1cm}$0.524$ \\
 2013 & \hspace*{0.1cm}$1,855$ & \hspace*{0.1cm}$54,976$ & \hspace*{0.1cm}$1,716$ & \hspace*{0.1cm}$25,685$ & \hspace*{0.1cm}$0.43$ & \hspace*{0.1cm}$5.80$ & \hspace*{0.1cm}$5.62$ &   \hspace*{0.1cm}$3.16$ &     [0.29, 5.74] & \hspace*{0.1cm}$0.738$ \\
 2014 & \hspace*{0.1cm}$2,206$ & \hspace*{0.1cm}$56,469$ & \hspace*{0.1cm}$2,037$ & \hspace*{0.1cm}$27,656$ & \hspace*{0.1cm}$0.41$ & \hspace*{0.1cm}$6.15$ & \hspace*{0.1cm}$5.57$ &  \hspace*{0.1cm}$10.42$ &    [7.64, 13.07] & \hspace*{0.1cm}$0.155$ \\
 2015 & \hspace*{0.1cm}$2,577$ & \hspace*{0.1cm}$57,276$ & \hspace*{0.1cm}$2,405$ & \hspace*{0.1cm}$30,590$ & \hspace*{0.1cm}$0.43$ & \hspace*{0.1cm}$6.27$ & \hspace*{0.1cm}$5.47$ &  \hspace*{0.1cm}$14.60$ &   [12.07, 17.05] & \hspace*{0.1cm}$0.043$ \\
 2016 & \hspace*{0.1cm}$3,021$ & \hspace*{0.1cm}$56,829$ & \hspace*{0.1cm}$2,796$ & \hspace*{0.1cm}$32,596$ & \hspace*{0.1cm}$0.41$ & \hspace*{0.1cm}$6.49$ & \hspace*{0.1cm}$5.34$ &  \hspace*{0.1cm}$21.39$ &   [18.92, 23.84] &                $<.001$ \\
 2017 & \hspace*{0.1cm}$3,613$ & \hspace*{0.1cm}$57,234$ & \hspace*{0.1cm}$3,384$ & \hspace*{0.1cm}$34,828$ & \hspace*{0.1cm}$0.42$ & \hspace*{0.1cm}$7.79$ & \hspace*{0.1cm}$5.77$ &  \hspace*{0.1cm}$35.02$ &   [32.52, 37.57] &                $<.001$ \\
 2018 & \hspace*{0.1cm}$4,436$ & \hspace*{0.1cm}$61,290$ & \hspace*{0.1cm}$4,106$ & \hspace*{0.1cm}$38,511$ & \hspace*{0.1cm}$0.42$ & \hspace*{0.1cm}$8.15$ & \hspace*{0.1cm}$6.22$ &  \hspace*{0.1cm}$30.93$ &   [28.73, 32.87] &                $<.001$ \\
 2019 & \hspace*{0.1cm}$5,382$ & \hspace*{0.1cm}$69,110$ & \hspace*{0.1cm}$5,036$ & \hspace*{0.1cm}$43,877$ & \hspace*{0.1cm}$0.41$ & \hspace*{0.1cm}$7.57$ & \hspace*{0.1cm}$5.61$ &  \hspace*{0.1cm}$34.87$ &   [32.76, 37.10] &                $<.001$ \\
\bottomrule
\end{tabular}

\end{center}
}}
\end{table}

\clearpage

\begin{table}[H]
{\fontsize{10}{10}\selectfont{
\caption{\textbf{Coarsened Exact Matching results for China-based papers produced in collaboration with the U.S.} $T$ and $C$ are the treatment and control populations, respectively; $T'$ and $C'$ are the populations of matched treatment and matched control papers, respectively; $\mathcal{L}_1$ is  the multivariate imbalance statistic~\cite{iacus2012causal}; $\mu_{T'}$ is the mean impact (i.e., mean $c_2$) of $T'$; $\mu_{C'}$ is the mean impact of $C'$; $\delta$ is the relative impact gain of $T'$ over $C'$, i.e., $\delta = 100\times( \mu_{T'} - \mu_{C'})/\mu_{C'}$; a bootstrap of 95\% confidence interval ($CI_{95\%}$) is provided; a t-test shows which $\delta$ values are statistically significant; see the resulting $p$-values. }

\label{supplementarytab:china_us_matching}
\begin{center}
\begin{tabular}{lcccccccccc}
\toprule
 year &                  $|T|$ &                   $|C|$ &                 $|T'|$ &                  $|C'|$ &        $\mathcal{L}_1$ &                 $\mu_{T'}$ &                $\mu_{C'}$ &                 $\delta$ &        $CI_{95\%}$ &                    $p$ \\
\midrule
  2005 &   \hspace*{0.1cm}$222$ &  \hspace*{0.1cm}$8,262$ &   \hspace*{0.1cm}$199$ &  \hspace*{0.1cm}$4,039$ & \hspace*{0.1cm}$0.37$ & \hspace*{0.1cm}$3.95$ & \hspace*{0.1cm}$1.85$ & \hspace*{0.1cm}$113.88$ &  [93.45, 130.29] & $<.001$ \\
 2006 &   \hspace*{0.1cm}$343$ & \hspace*{0.1cm}$13,607$ &   \hspace*{0.1cm}$319$ &  \hspace*{0.1cm}$7,094$ & \hspace*{0.1cm}$0.44$ & \hspace*{0.1cm}$3.11$ & \hspace*{0.1cm}$1.59$ &  \hspace*{0.1cm}$95.93$ &  [81.92, 109.73] & $<.001$ \\
 2007 &   \hspace*{0.1cm}$461$ & \hspace*{0.1cm}$13,857$ &   \hspace*{0.1cm}$434$ &  \hspace*{0.1cm}$7,404$ & \hspace*{0.1cm}$0.42$ & \hspace*{0.1cm}$3.67$ & \hspace*{0.1cm}$2.12$ &  \hspace*{0.1cm}$73.64$ &   [62.68, 83.32] & $<.001$ \\
 2008 &   \hspace*{0.1cm}$581$ & \hspace*{0.1cm}$20,437$ &   \hspace*{0.1cm}$547$ &  \hspace*{0.1cm}$9,647$ & \hspace*{0.1cm}$0.37$ & \hspace*{0.1cm}$3.90$ & \hspace*{0.1cm}$1.95$ & \hspace*{0.1cm}$100.14$ &  [90.24, 110.57] & $<.001$ \\
 2009 &   \hspace*{0.1cm}$770$ & \hspace*{0.1cm}$27,657$ &   \hspace*{0.1cm}$721$ & \hspace*{0.1cm}$14,979$ & \hspace*{0.1cm}$0.37$ & \hspace*{0.1cm}$3.82$ & \hspace*{0.1cm}$1.98$ &  \hspace*{0.1cm}$93.31$ &  [84.22, 101.51] & $<.001$ \\
 2010 &   \hspace*{0.1cm}$832$ & \hspace*{0.1cm}$33,184$ &   \hspace*{0.1cm}$776$ & \hspace*{0.1cm}$15,831$ & \hspace*{0.1cm}$0.36$ & \hspace*{0.1cm}$4.06$ & \hspace*{0.1cm}$1.97$ & \hspace*{0.1cm}$106.38$ &  [94.05, 115.64] & $<.001$ \\
 2011 & \hspace*{0.1cm}$1,042$ & \hspace*{0.1cm}$32,086$ &   \hspace*{0.1cm}$992$ & \hspace*{0.1cm}$16,629$ & \hspace*{0.1cm}$0.36$ & \hspace*{0.1cm}$4.39$ & \hspace*{0.1cm}$2.11$ & \hspace*{0.1cm}$107.75$ & [100.39, 115.50] & $<.001$ \\
 2012 & \hspace*{0.1cm}$1,209$ & \hspace*{0.1cm}$31,045$ & \hspace*{0.1cm}$1,147$ & \hspace*{0.1cm}$18,016$ & \hspace*{0.1cm}$0.35$ & \hspace*{0.1cm}$4.30$ & \hspace*{0.1cm}$2.58$ &  \hspace*{0.1cm}$66.17$ &   [61.16, 71.53] & $<.001$ \\
 2013 & \hspace*{0.1cm}$1,356$ & \hspace*{0.1cm}$31,093$ & \hspace*{0.1cm}$1,294$ & \hspace*{0.1cm}$18,698$ & \hspace*{0.1cm}$0.34$ & \hspace*{0.1cm}$4.99$ & \hspace*{0.1cm}$2.80$ &  \hspace*{0.1cm}$77.83$ &   [72.46, 83.26] & $<.001$ \\
 2014 & \hspace*{0.1cm}$1,770$ & \hspace*{0.1cm}$33,997$ & \hspace*{0.1cm}$1,671$ & \hspace*{0.1cm}$22,467$ & \hspace*{0.1cm}$0.34$ & \hspace*{0.1cm}$4.81$ & \hspace*{0.1cm}$3.11$ &  \hspace*{0.1cm}$54.60$ &   [50.68, 58.67] & $<.001$ \\
 2015 & \hspace*{0.1cm}$1,980$ & \hspace*{0.1cm}$31,635$ & \hspace*{0.1cm}$1,846$ & \hspace*{0.1cm}$21,733$ & \hspace*{0.1cm}$0.34$ & \hspace*{0.1cm}$5.72$ & \hspace*{0.1cm}$3.90$ &  \hspace*{0.1cm}$46.60$ &   [43.29, 50.16] & $<.001$ \\
 2016 & \hspace*{0.1cm}$2,339$ & \hspace*{0.1cm}$36,236$ & \hspace*{0.1cm}$2,213$ & \hspace*{0.1cm}$25,520$ & \hspace*{0.1cm}$0.34$ & \hspace*{0.1cm}$5.57$ & \hspace*{0.1cm}$3.86$ &  \hspace*{0.1cm}$44.18$ &   [40.97, 47.26] & $<.001$ \\
 2017 & \hspace*{0.1cm}$2,673$ & \hspace*{0.1cm}$41,097$ & \hspace*{0.1cm}$2,540$ & \hspace*{0.1cm}$29,483$ & \hspace*{0.1cm}$0.35$ & \hspace*{0.1cm}$6.27$ & \hspace*{0.1cm}$4.36$ &  \hspace*{0.1cm}$43.75$ &   [40.86, 46.34] & $<.001$ \\
 2018 & \hspace*{0.1cm}$3,402$ & \hspace*{0.1cm}$49,554$ & \hspace*{0.1cm}$3,215$ & \hspace*{0.1cm}$35,635$ & \hspace*{0.1cm}$0.34$ & \hspace*{0.1cm}$6.66$ & \hspace*{0.1cm}$4.97$ &  \hspace*{0.1cm}$33.88$ &   [31.63, 36.14] & $<.001$ \\
 2019 & \hspace*{0.1cm}$4,209$ & \hspace*{0.1cm}$64,023$ & \hspace*{0.1cm}$4,024$ & \hspace*{0.1cm}$48,540$ & \hspace*{0.1cm}$0.35$ & \hspace*{0.1cm}$6.20$ & \hspace*{0.1cm}$4.73$ &  \hspace*{0.1cm}$31.10$ &   [29.13, 33.03] & $<.001$ \\
\bottomrule
\end{tabular}

\end{center}
}}
\end{table}

\bibliography{main.bib}
\bibliographystyle{naturemag}